\documentclass[aps]{revtex4}
\usepackage{graphicx}
\usepackage{dcolumn}
\usepackage{bm}
\usepackage{rotate}
\usepackage{epsfig}
\newcommand \be  {\begin{equation}}
\newcommand \bea {\begin{eqnarray} \nonumber }
\newcommand \ee  {\end{equation}}
\newcommand \eea {\end{eqnarray}}
\begin{document}
\title{On a multi-timescale statistical feedback model for volatility fluctuations}
\author{Lisa Borland$^\dagger$}
\author{Jean-Philippe Bouchaud$^{+}$}
\email{bouchau@spec.saclay.cea.fr}
\affiliation{
$^{\dagger}$ Evnine-Vaughan Associates, Inc., 456 Montgomery Street, 
Suite 800, San Francisco, CA 94104,  USA\\
$^{+}$ 
Science \& Finance, Capital Fund Management, 6-8 Bd
Haussmann, 75009 Paris, France, and\\
Service de Physique de l'{\'E}tat Condens{\'e},
Orme des Merisiers,
CEA Saclay, 91191 Gif sur Yvette Cedex, France.\\
}
\date{\today}

\begin{abstract}
We study, both analytically and numerically, an ARCH-like, multiscale model of volatility, which 
assumes that the volatility is governed by the observed past price changes over different time scales. 
With a power-law distribution of time horizons, we obtain a model that captures most stylized facts of 
financial time 
series: Student-like distribution of returns with a power-law tail, long-memory of the volatility, 
slow convergence of the distribution
of returns towards the Gaussian distribution, multifractality and anomalous volatility relaxation 
after shocks. 
At variance with recent multifractal models that are strictly time reversal invariant, 
the model also reproduces the 
time asymmetry of financial time series: past large scale volatility influence 
future small scale volatility. In order to quantitatively reproduce all empirical observations, 
the parameters must be chosen such that the 
model is close to an instability, meaning that (a) the feedback effect 
is important and substantially increases the volatility, and (b) that the model 
is intrinsically difficult to calibrate
because of the very long range nature of the correlations. 
By imposing consistency of the model predictions 
with a large set of different empirical observations, 
a reasonable range of the parameters value can be determined. 
The model can easily be generalized to account for jumps, skewness and multiasset correlations. 
\end{abstract}

\maketitle

\section{Introduction}

The quest for a faithful mathematical model of price fluctuations has been 
taunting researchers for more than a
century now, starting with Bachelier's random walk model in 1900 \cite{Bachelier}. 
Such an endeavour is important 
for a bevy of reasons, both from the point of view of (a) 
fundamental economics (what is the cause of price variations and what 
information do they reveal?) and (b) of financial engineering, with option pricing, risk control and trading models as obvious applications. 

In an ideal world, ``the'' mathematical model of price changes should be simple 
enough to allow easy calculations and 
calibration, yet rich enough to embrace all known stylized facts that the 
recent access to huge amounts of 
data has helped establish. It is now widely accepted 
that price changes reveal (i) fat tails, well described by a
power-law decay of the probability distribution for large returns \cite{Longin,Lux,Gopi1}, 
(ii) long range memory in volatility fluctuations 
or volatility ``clustering'', again described 
by a power-law decay (in time) of the autocorrelation of 
the volatility \cite{volfluct1,volfluct2,Cizeau1} and (iii) asymmetric causal correlations 
between past price changes and future volatilities, 
often referred to as the ``leverage effect'' \cite{BMaP} (for reviews, 
see e.g. \cite{Guillaume,MS,Book,LuxR}). We discuss below 
other, somewhat related, stylized facts that have
 been reported in the recent literature, such as multifractal scaling, critical 
relaxation of the volatility after a shock (the financial analogue of the Omori law 
for earthquakes), etc.  More recently, 
some statistical {\it asymmetry} of financial time series under time reversal 
was pointed out \cite{LZ} -- in other words, 
financial time series do distinguish past from future. 
This might appear trivial but constitutes in fact, as we discuss below, 
a very strong constraint on the family of eligible models for financial time series 
-- for example, Bachelier's random
walk model is strictly time reversal symmetric.

Scores of different models have been proposed to improve upon 
the simple Brownian motion model, which has neither fat tails 
nor volatility clustering. L\'evy processes allow one to superimpose 
jumps to the Brownian motion, and therefore 
generate fat tails, but has no volatility clustering \cite{MS,Book,Levendorski,ContTankov}. 
GARCH models or simple stochastic volatility models such as the 
Heston model allow one to get both fat tails and some sort of volatility clustering, 
but not the long memory observed
in the data \cite{Heston,Fouque,Yakovenko,Perello}. Models that mix jumps and stochastic volatility 
have been investigated \cite{CGMY}.
Multifractal stochastic volatility models, initiated by Mandelbrot, Fisher and Calvet 
\cite{MCF} and much studied since 
\cite{Ghasg,multif1,multif2,CF1,CF2,BPM,BMD1,BMD2,Lux1,Lux2,MandelB,Pochart,kertecz}, 
seem to capture in a parsimonious way a large amount of empirical properties. 
However, most multifractal models are again
strictly time reversal symmetric and lack an intuitive interpretation in 
terms of agent based trading models \cite{BBMZ}. 
We in fact strongly 
believe that any serious model of price fluctuations should {\it in fine} be 
justified by reasonable behavioral rules
and market microstructure effects (see \cite{LuxM,minority,BrockH,Chiarella,GB} 
for recent work in that direction). Quite recently, one of us (LB) has proposed, 
in the context of option pricing, a 
``statistical feedback'' process where the local volatility is large when price 
moves are deemed rare, leading to 
a non-linear diffusion equation for the price \cite{LB1}. 
This equation can be solved and leads to a Student-Tsallis distribution 
for price changes at all times \cite{Tsallis}. In its original form, however, the model 
breaks time translation symmetry: there is a 
well defined starting date and starting price. Although this can be used to price options \cite{LB1,LBB} 
(in the spirit of the Hull-White model for interest rates \cite{Hull}), the process has to be modified to be interpreted 
as a {\it bona fide} model of returns. Such an extension, and its modification to account for long-range memory, was
proposed in \cite{LBnew} and recovers, following a different route, a multiscale GARCH model proposed by Zumbach and 
Lynch in 2003 \cite{LZ,Zumb2} (see \cite{volfluct2,FIGARCH,HARCH} for earlier work in that direction). 
Numerical simulations of this model suggest a very rich phenomenology, that seems to account 
for most stylized facts of financial time series.

The aim of the present paper is to motivate this new model, discuss its relation with previous work, and 
investigate in full details its statistical properties, both analytically and numerically. We focus in particular on 
the probability distribution of returns which is the crucial ingredient for option pricing and risk control. Although
not an exact result, we find that these distributions can be well fitted by a Student-Tsallis form, with a lag-dependent 
tail exponent. We reproduce in great details most empirical facts, including the anomalous relaxation of the volatility 
after a shock, and the past/future asymmetry of the time series. The model can be generalized to include jumps, 
the leverage effect, and multi-stock correlations. We then discuss the issue of calibration. Within strict econometric 
standards, calibration is extremely difficult due to the long-memory nature of both the empirical 
volatility process and the theoretical models that are constructed precisely to capture this long memory. We advocate the
idea of `soft' calibration, which in such cases should consist in reproducing semi-quantitatively as many observables 
as possible. These observables should be chosen to be robust to the details of the model specification, and test different 
``orthogonal'' predictions of the model (these statements will be made clearer in the course 
of the paper and in 
 Section \ref{calibration}). Consequences for option pricing are briefly discussed, and will be the subject of another paper.

\section{Set up and motivation of the model}
\label{themodel}

In the following, we will consider a discrete time model, with an elementary time scale equal to $\tau$, for example
$\tau=$1 minute. [A continuous time version of the model will be discussed below].
The price at time $t_i=i \tau$ will be noted $p_i$. We will conform to the standard of dealing 
with the log-price 
$x_i=\ln p_i$ and define returns as $r_i=x_{i+1}-x_i$.\footnote{Note however that a model based on log-prices
has no deep theoretical justifications and might not be the best representation of reality. 
See \cite{Book,BMaP} for a detailed discussion of this point.}
The random return is constructed as the product of a time dependent volatility $\sigma_i$ and a 
random variable $\xi_i$ of zero mean and unit variance:
\be\label{return}
r_i = \mu \tau + \sigma_i \xi_i \sqrt{\tau},
\ee
where $\mu$ is the average drift, which we will set to zero in the sequel, 
meaning that we measure all returns relative
to the average drift. The noise $\xi_i$ can a priori have any
probability distribution to account for high frequency kurtosis and jumps, 
but for simplicity we will mostly focus 
in this paper on the case of a Gaussian noise. 
However, as we discuss below, the introduction of jumps is needed to 
faithfully reproduce real price time series. 

The seminal insight of ARCH or GARCH models \cite{reviewarch}
is that the volatility process reflects trading activity and is 
subordinated to past price changes. Intuitively, 
the level of activity becomes high when past price changes are, in some sense, anomalous. In the simplest ARCH model,
this is expressed as:
\be\label{arch}
\sigma_i^2 = \sigma_0^2 \left[1 + g \frac{r_{i-1}^2}{\sigma_0^2 \tau}\right],
\ee
meaning that the volatility is equal to its `base level' $\sigma_0^2$ plus a contribution coming from the last 
price change. In fact, we have written the feedback term in a way that expresses the comparison between the 
square of the last return and its expected value, equal to $\sigma_0^2 \tau$. If the last price change was small 
compared to usual, the volatility today is close to its normal value, whereas in the other limit, the last return is
deemed anomalous and leads to a potentially large increase of today's activity. 

\begin{figure}
\begin{center}
\psfig{file=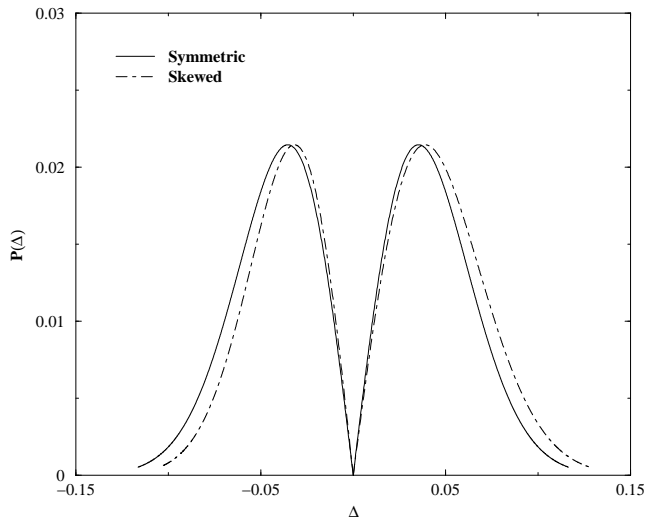,width=7cm,angle=270} 
\end{center}
\caption{Schematic shape of the distribution of stop loss/stop gain thresholds 
around the opening price of the
trade. Plain line: symmetric distribution; dotted line: 
asymmetric distribution, giving rise to the leverage effect
discussed in section \ref{leverage}.}
\label{Fig1}
\end{figure}

An argument motivating Eq. (\ref{arch}) 
above is as follows. Suppose that some traders open positions (for example, long) at time $t_{i-1}$, when 
the price is $p_{i-1}$. Such trades
are often initiated with both a profit objective and a risk limit, 
which would close the position at time $t_i$ if the
price has moved up too much (stop gain) or down too much (stop loss). 
It is very natural to hypothesize that to each 
opening trade are associated two thresholds, one above, one below $p_{i-1}$, 
that trigger a closing trade if exceeded. If
many agents open both long and short trades at $t_{i-1}$, one can expect a quasi-continuous distribution 
of 
thresholds at $p_{i-1}(1+\Delta)$, 
more or less symmetrically distributed around $p_{i-1}$, triggering with 
equal probability sell back or buy back orders. 
The density $P(\Delta)$ is, in the simplest case, even (but see 
Section \ref{leverage} for the inclusion of the leverage effect) and obviously vanishes at $\Delta=0$ 
since nobody opens a trade to close it immediately (see Fig. 1). The width of $P(\Delta)$ is 
given, in order of magnitude, by $\sigma_0 \sqrt{\tau}$ since this gives 
the natural scale beyond which 
an event might be deemed anomalous. 
Hence, a relative change of price $r_{i-1}$ will trigger on the order of:
\footnote{The following argument is only intended to be qualitative. 
In reality, the trading range between 
$t_{i-1}$ and $t_i$ is given by the high minus low of that period, 
rather than the open to close; all thresholds 
with that interval will be hit. However, the high-low is of the same order as the 
open-close $r_{i-1}$ -- in fact,
for a random walk, the average high-low is exactly twice the average $|r_{i-1}|$, 
a result due to Bachelier himself!}
\be
N_i(r_{i-1}) \approx \int_0^{|r_{i-1}|} P(\Delta) d\Delta 
\ee
stop trades. These trades of random sign lead, on the next day, 
to an increase of the volatility as:
\be
\sigma_i^2 = \sigma_0^2 + G N_i/\tau,
\ee
where $G$ is the average square impact per trade, 
and $\sigma_0^2$ is the volatility due to all other trades. 
Taking into account that $P(\Delta)$ extends over a 
range $\sigma_0 \sqrt{\tau}$, one finally obtains a general
single time scale ARCH model:
\be\label{g-arch}
\sigma_i^2 = \sigma_0^2 \left[1 + {\cal G}(\frac{|r_{i-1}|}{\sigma_0 \sqrt{\tau}})\right],
\ee
where the function $\cal G$ depends on the detailed shape of $P(\Delta)$. Taking for simplicity, in 
accordance with the above discussion, 
\be
P(\Delta) = P_1 \frac{|\Delta|}{2\beta^2 \sigma_0^2 \tau} 
\exp(-\frac{\Delta^2}{2\beta^2 \sigma_0^2 \tau}),
\ee
(where $\beta$ is a number setting the width of the distribution of thresholds, 
and $P_1$ the total number of 
opened trades) finally leads to:
\be
{\cal G}(u) = 2g \beta^2 \left(1 - \exp(-u^2/2\beta^2)\right),
\ee
where $g={GP_1}/{2\beta^2 \sigma_0^2 \tau}$ is the ratio measuring the impact of all stop trades 
compared to that of all other trades. The simplest ARCH model Eq. (\ref{arch}) corresponds to the 
limit $u \ll \beta$,
that is, neglects saturation effects related to the fact that stop limits are not placed 
arbitrarily far from the 
entry point (i.e. $\beta$ is finite). When this saturation is neglected, ${\cal G}(u)$ 
is simply given by $gu^2$ (but see below, Fig. 11).

Although the above feedback mechanism is most probably at play in financial markets, 
a strong limitation of the above model 
is to consider that all traders have the same time horizon, equal to $\tau$ in the above formulation. 
However, it
is well documented that the activity of financial markets is 
fueled by traders with different time horizons, from a
few hours to a few months or even years (see e.g. \cite{HARCH,LZ}). 
Therefore, stop losses or profit objectives are not placed only around the
last price $p_{i-1}$ but around possibly all past prices $p_{i-\ell}$, $\ell=1,2,...$. Correspondingly,
the width of the distribution of these thresholds is calibrated to the 
volatility of the price over the particular 
trading horizon, i.e. $\sigma_0 \sqrt{\ell \tau}$.  
The generalization of Eq. (\ref{g-arch}) to this situation 
therefore reads:
\be\label{g-arch-bis}
\sigma_i^2 = \sigma_0^2 \left[1 + \sum_{\ell=1}^\infty 
{\cal G}_\ell(\frac{|x_i-x_{i-\ell}|}{\sigma_0 \sqrt{\ell \tau}})\right].
\ee
Expanding ${\cal G}_\ell$ for small arguments finally leads to the symmetric version of the 
model studied in the present paper (the inclusion of asymmetry will be discussed in Section \ref{leverage}):
\footnote{It might actually be more consistent, at the level of the theoretical justification of the
model, to replace $\sigma_0^2$ in the feedback term by a moving average of the past volatility itself,
since traders are sensitive to the recent level of volatility.
This would make the model more complicated, and we leave this extension for future investigation.}
\be\label{the-model}
\sigma_i^2 = \sigma_0^2 \left[1 + \sum_{\ell=1}^\infty 
g_\ell \frac{(x_i-x_{i-\ell})^2}{\sigma_0^2 \ell \tau}\right],
\ee
with
\be
r_i= x_{i+1}-x_i =\sigma_i \xi_i \sqrt{\tau}.
\ee
The coupling constant $g_\ell$ is proportional to the number of trades $P_\ell$ 
with horizon $\ell$. Because
traders with a longer horizon have slower trading frequencies and 
under-react compared to short term traders, it 
is reasonable to imagine that $g_\ell$ is a decaying function of $\ell$. 
Both for simplicity and because it allows 
us to reproduce several stylized empirical facts, we will choose $g_\ell$ to be an inverse power:
\be 
g_\ell = g/\ell^\alpha,
\ee   
but other choices are possible. For example, Zumbach and Lynch have presented evidence 
that $g_\ell$ has additional peaks on the day, 
week and month times scales. These authors have proposed a model 
very close in spirit to Eq. (\ref{the-model}), and discussed some of its properties. 
In fact, Eq. (\ref{the-model}) 
is a special case in the family of quadratic ARCH models, where the 
volatility is expressed as a general quadratic form of past returns:
\be\label{q-arch}
\sigma_i^2 = \sigma_0^2 + \sum_{j<i, k<i} {\cal M}(i;j,k) \frac{r_j r_k}{\tau},
\ee
which contains ARCH, GARCH, etc. 
Our specification insists that only combinations of returns `reconstructing' 
actual price changes over different time scales occur in the above sum, 
because they correspond to quantities
directly observable to the crowd of traders, which, we argue, 
strongly influence the trading at time $i$. Our model
corresponds to a particular choice for $\cal M$ above:
\be
{\cal M}(i;j,k) = \sum_{\ell=\max(i-j,i-k)}^\infty \frac{g_\ell}{\ell},
\ee
whereas most ARCH models correspond a certain 
regression on past {\it instantaneous} square returns, i.e., to 
${\cal M}(i;j,k) = K(i-j) \delta_{jk}$, with a certain kernel function $K$,
usually corresponding to an exponential moving average, $K(\ell)=\alpha^\ell$. 

With a power-law specification for $g_\ell$, and the choice of a Gaussian distribution 
for the noise term $\xi$ in the definition of returns, our model is fully 
determined by only four parameters: $\sigma_0$ sets the volatility scale, 
$\tau$ sets the shortest time scale
over which feedback effects are effective, $g$ measures the strength 
of these feedback effects and $\alpha$ describes the relative importance 
of short term traders and long time traders in the
feedback process. It may however be that the assumption of a Gaussian noise
for $\xi$ is insufficient to account for the high frequency statistics of the returns. In particular,
one expects that true `jumps' related to unexpected news are not described in terms of a volatility 
feedback process. It is easy to extend the model in that direction and choose another distribution
for $\xi$. In the following sections, we will present several 
analytical and numerical results of the Gaussian version of this
model, and compare them to empirically known results. But before doing so, let us give the 
continuous time formulation of the same model, which can be convenient for some applications, such
as option pricing. Introducing the standard Brownian noise $dW_t$, one may write:
\be
dx_t = \sigma_t dW_t,
\ee
with:
\be
\sigma_t^2 = \sigma_0^2 + 
g \tau^\alpha \int_{-\infty}^t {\rm d}t' \frac{(x_t - x_{t'})^2}{(t - t' + \tau)^{1+\alpha}}.
\ee
This model is well defined as soon as $\alpha > 1$, which is the case we will focus on in the sequel.

\section{Analytical results}
\label{analytics}

\subsection{Unconditional distribution of the volatility}

Although our model (Eq. (\ref{the-model})) expresses the volatility as a deterministic 
function of the past prices
and the only source of randomness comes from the noise $\xi_i$ in Eq. (\ref{return}), 
the volatility effectively 
appears as a random variable, and one can ask questions about its distribution, correlations, etc. The 
simplest question concerns the average value of the volatility, which also coincides, for a 
stationary process, with the 
long term volatility of the price. 
Averaging will always be denoted below with brackets $\langle \dots \rangle$ 
around the quantity which is averaged. For the average volatility, one has (assuming stationarity) :
\be
\langle \sigma^2 \rangle = \langle \sigma_i^2 \rangle = \sigma_0^2 + \sum_{\ell=1}^\infty 
g_\ell \frac{\langle (x_i-x_{i-\ell})^2 \rangle}{\ell \tau} =
\sigma_0^2 + [\sum_{\ell=1}^\infty g_\ell] \langle \sigma^2 \rangle.
\ee
This equation has a well behaved solution only if:
\be
z_2 = \sum_{\ell=1}^\infty g_\ell < 1,
\ee 
where the above equation defines $z_2$, the subscript `$2$' refers to the fact that we study
here the second moment of the volatility. When $z_2 < 1$, the square
volatility is amplified by a factor $1/(1-z_2)$ compared to the initial value $\sigma_0^2$. 
In the case $z_2 \geq 1$, on
the other hand, the process becomes non stationary and 
the volatility grows without bound as time elapses. It is 
clear that the condition $z_2 < 1$ can only be met if the sum of 
$g_\ell$ converges, which imposes that the exponent
$\alpha$ is larger than one. 
For $\alpha > 1$, one finds $z_2 = g \zeta(\alpha)$, which delimits a region in the
plane $g, \alpha$ where the process is stationary. 
In the following, we will often assume that $\alpha$ is larger than 
unity but close
to it (which is suggested by empirical data), and use in this limit a 
continuous approximation for discrete 
sums. In particular, $\zeta(\alpha) \approx 1/(\alpha-1)$. 
We will find below that empirical data on stocks favors 
values of $z_2 \approx 0.85 - 0.9$, meaning that the square volatility is increased by a 
factor $\sim 6-10$ compared to
its initial value $\sigma_0^2$. Therefore feedback effects might be an 
important cause of the excess volatility in financial markets \cite{Shiller,WB}.

In order to compute higher moments of the volatility, one needs in general 
to know the full temporal correlation 
of the volatility, that we will establish in the next paragraph. 
Simplified, approximate calculations 
can be performed in two extreme cases: 
(i) no temporal correlations (ii) full temporal correlations. This leads to
an equation for $\langle \sigma^4 \rangle$ of the form 
$(1-z_4)\langle \sigma^4 \rangle=${\sc rhs}, where
the right hand side is finite whenever $\alpha > 1$, and can be 
computed if necessary (see below). The important 
discussion concerns the value of $z_4$. 
We will denote $M_{k}={\cal M}(0;-k,-k)$, which behaves, for large $k$, 
as $k^{-\alpha}/(\alpha-1)$. Using the results established below (see Eq. (\ref{full})),
one can obtain a lower bound $z_{4,<}$ and an upper bound $z_{4,>}$ on the value of $z_4$.
If correlations are neglected, one finds:
\be
z_4 \geq z_{4,<} = 3g^2 \sum_{k=1}^\infty M_{k}^2.
\ee
If on the other hand, if correlations are overestimated and taken to be constant in
time, one finds an upper bound for $z_4$ that reads:
\be
z_4 \leq z_{4,>} = g^2 \left([\sum_{k=1}^\infty M_{k}]^2
+ 2 \sum_{k=1}^\infty M_{k}^2 + 4 \sum_{k=1}^\infty (k-1) M_{k}^2 \right).
\ee
As long as $z_4 < 1$, the fourth moment of $\sigma$ is finite, but if $z_4$ reaches unity, 
it does diverge, leading to
an infinite kurtosis for the returns. For $\alpha=1.15$, 
we find $z_{4,<}=0.16 \, z_2^2$ and 
$z_{4,>}=1.44 \, z_2^2$. This shows that the kurtosis $\kappa$ is certainly 
finite for $z_2 < 0.833$; numerical simulations below 
suggest that $\kappa$ indeed remains finite beyond that value. 

The above argument is easily generalized to higher even 
moments of $\sigma$, leading to an equation 
$(1-z_{2n})\langle \sigma^{2n} \rangle=${\sc rhs} with:
\be
z_{2n,<} = (2n-1)!!\, g^n \sum_{k=1}^\infty M_{k}^n,
\ee
and a more cumbersome expression for $z_{2n,>}$. 
For large $n$, one finds $z_{2n,<} \sim (2g n/e)^n$, showing that 
however small the value of $g$, sufficiently 
high moments of the volatility are divergent. 
Since  $r_i= \sigma_i \xi_i$, the even moments of the returns are 
given by:
\be
\langle r^{2n} \rangle = (2n-1)!! \langle \sigma^{2n} \rangle;
\ee
therefore high moments of returns themselves diverge, 
suggesting that both the unconditional distribution of volatility 
and returns have a power-law tail (possibly multiplied by a slow function), 
with an exponent equal to the order of the last finite moment. We will confirm this
prediction numerically in the following section.
Remember however that the above discussion is only valid when the noise $\xi$ is Gaussian; 
if $\xi$ itself
has a non zero kurtosis, 
then its contribution should be taken into account. 

\subsection{Temporal correlations of the volatility}

A well known stylized fact is that the volatility is a `long-memory' process, 
which means that the temporal 
correlations of the square volatility decay as an inverse power of the time lag, 
$\ell^{-\nu}$, with an exponent $\nu$ less than unity.
This property turns out to be extremely important because 
it is at the root of the very slow convergence of
the distribution of aggregated returns towards the Gaussian. 
More precisely, the kurtosis of the return $x_i-x_{i-\ell}$ 
over scale $\ell$, itself decays as $\ell^{-\nu}$ instead of $\ell^{-1}$, 
which is the case when the volatility process has a short 
memory. Since the empirical value of $\nu$ is, for stocks, 
on the order of $\nu=0.2-0.3$, the slowing down is
substantial and essential to explain why long dated options still have a smile.

We therefore turn to the calculation of the correlation function of the volatility, defined as:
\be
{\cal F}(\ell)=\frac{\langle \sigma_{i+\ell}^2 \sigma_i^2 \rangle}{\langle \sigma^2 \rangle^2} - 1.
\ee
In the limit $g^2 \ll 1$, one can quite easily perform a perturbative analysis 
that neglects terms of order $g^4$, to
get:
\be\label{Gsecondorder}
{\cal F}(\ell) = 2 g^2 \left[\sum_{0<k<j} \frac{k^{1-\alpha}}{(\ell+j)^{1+\alpha}} +
\sum_{0 < j \leq k} \frac{j^2}{k^{1+\alpha} (\ell+j)^{1+\alpha}} \right].
\ee
An analysis of this result for $\ell \gg 1$ finally gives, for $\alpha > 1$ 
but close enough to unity such that
one can use continuous integrals instead of discrete sums:
\be\label{Gasymptotic}
{\cal F}(\ell) \sim \frac{4 g^2 \Gamma(2-\alpha) 
\Gamma(2\alpha-1)}{\alpha^2 \Gamma(\alpha)} \ell^{2-2\alpha} 
\equiv {\cal F}_\infty \ell^{-\nu},
\ee
leading to a kurtosis exponent $\nu=2\alpha-2$. 
The volatility is a long memory process whenever $\nu \leq 1$, i.e.
$1 < \alpha \leq 3/2$. Comparison with empirical data, done below, 
suggests that $\alpha$ is in the range $1.1 - 1.2$. 
The exact equation for ${\cal F}(\ell)$, not restricted to small $g^2$, 
can also be written down, although it
is more cumbersome. For this calculation, one should note that 
averages such as $\langle \sigma_i^2 \xi_j^2 \rangle_c$ 
(where the subscript $c$ denotes a connected average) are 
non trivial, since the volatility randomness comes entirely 
from past returns themselves. This contrasts with many
stochastic volatility models where the volatility $\sigma_i$ and the noise $\xi_j$ 
are often chosen to be independent
(unless one wants to model the leverage effect). In the present case, one finds, for $j< i$:
\be
\langle \sigma_i^2 \xi_j^2 \rangle_c = g^2 \sum_{k,k' > 0} {\cal M}(0;-k,-k') 
\langle \sigma_{i-k} \sigma_{i-k'} \xi_{i-k} \xi_{i-k'} \xi_j^2 \rangle_c = 
2g^2 \langle \sigma^2 \rangle M_{i-j},
\ee
Now, the full self-consistent equation for $\cal F$ reads:
\bea\label{full}
{\cal F}(\ell) &=& g^2 [3 {\cal F}(0) + 2] \sum_{k=1}^\infty M_{k} M_{k+\ell} + 
4g^2 \sum_{k>k'=1}^\infty M_{k} M_{k+\ell} [1+{\cal F}(k-k')+2g^2 M_{k-k'}] \\
&+&2g^2 \sum_{k>k'=1}^\infty M_{k} M_{k'+\ell} [{\cal F}(k-k')+2g^2 M_{k-k'}] + 
g^2 \sum_{k=1}^\infty \sum_{k'=1}^\ell M_k M_{k'} [{\cal F}(\ell-k'+k) + 2g^2 M_{\ell-k'+k}]
\eea
Specializing to $\ell=0$ leads to the fourth moment 
of the volatility studied in the above section. The 
two assumptions made there to obtain a lower and an upper bound correspond to 
${\cal F}(\ell)={\cal F}(0) \delta_{\ell,0}$ and 
${\cal F}(\ell)={\cal F}(0)$, respectively. 
For large $\ell$, an asymptotic estimate of the various terms
leads to the same decay as that predicted by the above perturbative calculation, 
i.e., ${\cal F}(\ell) \sim
{\cal F}_\infty \ell^{-\nu}$, with $\nu=2\alpha-2$, 
and a prefactor ${\cal F}_\infty$ increased by a 
factor $1/(1-z_2^2)$. However, sub-dominant terms also appear, 
proportional to $\ell^{-2\nu}$, $\ell^{-\alpha}$, etc.
The finite $\ell$ behaviour of ${\cal F}(\ell)$ would require to solve the above equation 
numerically. 

From the knowledge of ${\cal F}(\ell)$ one can obtain the $\ell$ dependence 
of the kurtosis of the returns, following 
\cite{Book}. Again, one should take care of the terms involving 
$\langle \sigma_i^2 \xi_j^2 \rangle_c$, which, as 
we discuss below, lead to a new, perhaps unexpected effect. 
One finds, for the kurtosis of the returns on lag $\ell$: 
\be\label{kurtosis}
\kappa(\ell) = \frac{1}{\ell} \left[\kappa(1) + 
6 \sum_{j=1}^\ell (1-\frac{j}{\ell}) [{\cal F}(j)+ 2g^2 M_j ] \right].
\ee
For large lags $\ell \gg 1$, one finds, 
using Eq. (\ref{Gasymptotic}), and for $\alpha$ close to $1$:
\be
\kappa(\ell) \sim \frac{3 {\cal F}_\infty}{(3-2 \alpha)(2-\alpha)} \ell^{-\nu}.
\ee
Therefore, one expects the returns to converge to Gaussian, 
but only on a very long time scale. Any measure of the
distance from a Gaussian -- such as the mean absolute moment studied below -- 
will tend to zero very slowly, as 
$\ell^{-\nu}$, see Figs 4-a, 4-b. 
If one now studies Eq. (\ref{kurtosis}) for small values of $\ell$, say $\ell=2$, one finds:
\be
\kappa(2)-\kappa(1)=3 [{\cal F}(1)-{\cal F}(0)+ 2 g^2 M_1]
\ee
In many models, the last term is absent, and since 
${\cal F}(1) \leq {\cal F}(0)$, one usually finds that the kurtosis
of aggregated returns is less than the kurtosis of elementary returns. However, the third term in
the above expression suggests that one can observe, in some cases, a kurtosis that 
first {\it increases} with
lag before decaying to zero. 
We will see that this is indeed the case in the numerical simulations of our model,
although this effect is, again, very sensitive to the assumption that $\xi_i$ is a purely Gaussian noise.

\subsection{Conclusion}

The summary of this technical section is that the two major stylized facts 
(fat tails and volatility long-memory)
are present in our model. We have indeed shown 
that the distribution of returns and of the volatility have power-law
like tails, since high moments of these distributions diverge. 
We have also shown that the temporal correlation 
of the volatility is decaying as a slow power law. 
The following sections will be to establish these properties
more quantitatively using numerical simulations, 
and to show that many more stylized facts can be reproduced by 
the model. Finally, we will turn to the question of calibration 
and discuss how the model parameters can be
chosen to fit empirical data.  

\section{Numerical results}
\label{numerics}

We have established above that the volatility-volatility correlation function, 
and the kurtosis, decay at long times 
as $\ell^{-\nu}$ with $\nu=2(\alpha-1)$. 
A large amount of empirical work on financial time series suggest that
$\nu$ is in the range $0.2 - 0.4$ for many different assets. 
For example, averaging over the 500 largest 
stocks of the NYSE leads to $\nu \approx 0.25$, 
while $\nu \approx 0.3$ for the S\&P 500 Index \cite{Book}. 
We therefore choose to fix $\alpha=1.15$ (corresponding to $\nu=0.30$) 
in most of the numerical experiments that
we have conducted. Other values of $\alpha$ are briefly discussed, 
in particular in the context of the model calibration.
The choice $\alpha=1.15$, although guided by empirical data, 
immediately leads to a numerical problem due to its
proximity with the critical value $\alpha=1$ which separates a (theoretically) 
stationary regime for $\alpha > 1$
from a non stationary regime for $\alpha \leq 1$. 
The convergence of (say) the average volatility to its asymptotic
value is expected to occur at speed $T^{1-\alpha}$, 
where $T$ is the total length of the time series. For $\alpha=1.15$,
this is extremely slow: even for $T=10^6 \tau$, one expects corrections 
of order $10 \%$ to the theoretical asymptotic
results. For this reason, and also to speed up the numerical calculation of 
the sum that determines the volatility 
(Eq. \ref{the-model}), we have truncated the power-law memory kernel 
$g_\ell$ beyond $\ell=5.\ 10^4$. The total
length of our simulations is usually $10^6$ steps, 
but we discard the first $15.\, 10^4$ points of the series 
before we start measuring any observable. Although this is, again, 
insufficient to obtain very precise 
results for such low values of $\alpha$, we believe that these numerical experiments 
are sufficient to obtain 
a good estimate of a host of different interesting observables, 
in any case comparable in quality to the 
corresponding estimates on real price time series. 
As will be clear below, we estimate that one day 
corresponds in our model to $\ell \sim 300$; therefore $10^6$ time steps 
corresponds to $3000$ trading days, or twelve
years of data. In the following, the base volatility $\sigma_0$ is set to 
$\sigma_0=1$, any other value 
would only change the following results by a trivial multiplicative factor on the returns. 
We will vary the 
coupling constant $g$, which we will in fact express in terms of  $z_2= \sum_\ell g_\ell$, 
since we know from the
above discussion in section \ref{analytics} that it is really $z_2$ that 
measures the strength of feedback 
effects on the volatility. In the limit $z_2 \to 1$, we know from section \ref{analytics} 
that the volatility will blow up
and the process becomes non-stationary for all values of $\alpha$. 
Therefore, studying numerically values of 
$z_2$ too close to unity will also be difficult (the convergence is now as
slow as $[(1-z_2)T]^{1-\alpha}$!), but, ironically, corresponds 
to the empirical situation. 
In the following, we restrict our simulations to the range $z_2 \in [0.60,0.85]$ --
smaller values of $z_2$ lead to a process which is only weakly non Gaussian, 
whereas larger values of $z_2$ 
give rise to a numerically very unstable process, 
even though in theory the process should still be stationary on
extremely long time scales. We will see below that values of $z_2$ as high as
$0.9$ might be needed to fit the data, but we have not attempted to simulate the 
model for such a large value.

Although the issue of calibration will be more deeply discussed in section \ref{calibration}, 
we will compare in this
section our numerical results to empirical data, averaged over a set of 252 US stocks, 
chosen among the most liquid ones, 
during a four year time period: 2000-2003. 

\begin{figure}
\begin{center}
\psfig{file=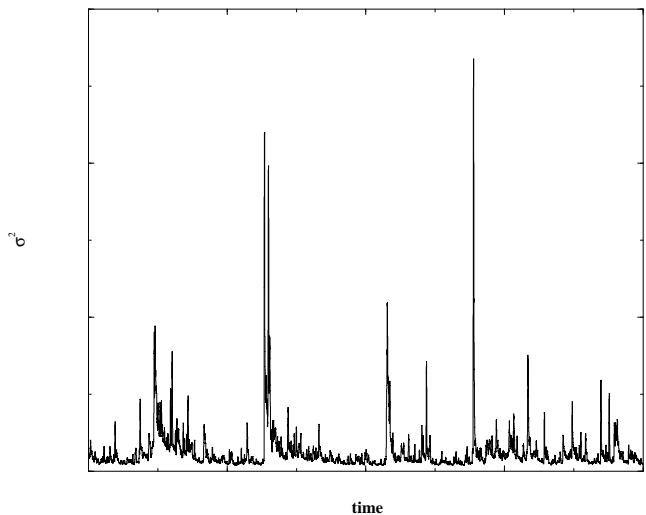,width=7cm,angle=270} 
\end{center}
\caption{A typical time series (of length $2\, 10^5$) of the 
volatility $\sigma^2$,  
for $z_2=0.85$ and $\alpha=1.15$. 
We have in fact shown a 300 $\tau$ moving average of $\sigma_i^2$, 
aimed at representing the `daily' volatility within our model.}
\label{Fig2}
\end{figure}

\subsection{Volatility distribution and volatility correlations}

\subsubsection{Volatility distribution}

We first focus on the properties of the `true' volatility $\sigma_i$, 
which we can of course measure numerically
but is unobservable directly in practice: only proxies of the volatility, 
obtained by averaging over several time 
steps, can be studied. 
A typical time series of $\sigma^2$ is shown in Fig. 2, and reveals apparent 
shocks and volatility clustering 
familiar in financial time series. 
We show in Fig. 3 the histogram of $u=\ln \sigma$ for different values of $z_2$. 
Obviously, since $\sigma > \sigma_0=1$, the probability distribution function (pdf) 
of $u$ is zero when $u \leq 0$. 
We have found that the pdf $P(u)$ of $u$ can be very accurately 
fitted by the following form (see Fig. 3):
\be\label{beta}
P(u) = Z \exp\left[-(\frac{u_0}{u})^\beta - \mu u\right] \Theta(u),
\ee
where $\Theta(u>0)=1$ and $\Theta(u < 0)=0$.
We have no detailed justification for this specific functional form for $u \to 0$. 
On the other hand, it is easy to show 
that the exponential tail for large positive $u$ 
translates into a power-law distribution for $\sigma$
itself, decaying as $\sigma^{-1-\mu}$, which is indeed expected from our theoretical analysis. 
Correspondingly, the distribution of returns will also display the same power-law tail. 
The values of $\mu$ that we find using the 
above fit are summarized in Table I. From Fig. 3, however,
we see that the apparent slope of $\ln P(u)$ vs. $u$ in the available
range of `large' $u$ values 
is slightly smaller than the value of $\mu$ obtained 
from a global fit with Eq. (\ref{beta}). 
For example, for $z_2=0.85$, we find $\mu \approx 4$, but the 
apparent slope is $\approx 3.5$, interestingly 
closer to the value reported for stocks 
$\mu \approx 3$ \cite{Gopi1}. A slightly larger
value of $z_2 = 0.9$ would be in even better agreement 
with this empirical value of the exponent (see Table I). 

\begin{figure}
\begin{center}
\psfig{file=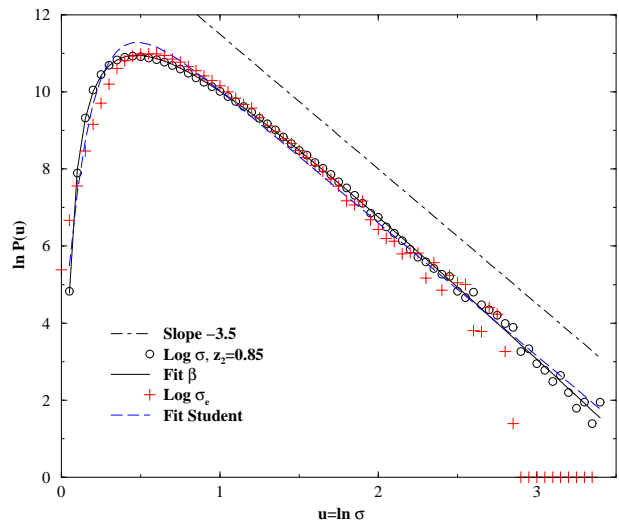,width=7cm,angle=270} 
\end{center}
\caption{Histogram of $u=\ln \sigma$ for $z_2=0.85$ and $\alpha=1.15$, and two fits, 
using Eq. (\protect\ref{beta}) -- ``Fit $\beta$'' --
and Eq. (\protect\ref{inv-Gamma}) -- ``Fit Student''. 
We also show the slope $\mu=-3.5$ for comparison with the tail of the distribution
of $u$. The pluses correspond to the histogram of the average volatility over $100$ 
time steps, close to what one would determine empirically from price time series.}
\label{Fig3}
\end{figure}

We have also tried to fit $P(u)$ assuming an inverse Gamma distribution for the pdf of 
$\sigma$ \cite{Mantegna,Book}, which corresponds to a Student 
distribution for the returns. In terms of $u=\ln \sigma$, this reads:
\be\label{inv-Gamma}
P(u) = Z' \exp\left[-A e^{-Bu} - \mu u\right],
\ee
where $A$ and $B$ are parameters, and $\mu$ is the power-law tail of the
return distribution. 
Of course, this distribution cannot be exact 
in the present case since it takes non zero values when 
$u < 0$. Although it is definitely not as good a fit as Eq. (\ref{beta}), it is quite
acceptable, meaning that returns are indeed close to being Student distributed in our model. 
On the other hand, a log-normal distribution for $\sigma$ is clearly inadequate to describe our data
(it would correspond to a parabola in Fig. 3.)

\begin{table}
\begin{center}
\begin{tabular}{||l|c|c|c|c|c|c|c|c|c||} \hline\hline
$z_2$ \ \hspace{0.1cm} \   & 
\hspace{0.1cm}  $(1-z_2)^{-1}$ \hspace{0.1cm} &
\hspace{0.1cm}  $\langle \sigma^2 \rangle$ \hspace{0.1cm} &
\hspace{0.1cm}  ${\cal F}(0)$ \hspace{0.1cm} &
\hspace{0.2cm}  $\mu$ \hspace{0.2cm} &
\hspace{0.2cm}  $\beta$ \hspace{0.2cm} &
\hspace{0.2cm}  $u_0$ \hspace{0.2cm} &
\hspace{0.1cm}  $\lambda^2$ ($\ln$) \hspace{0.1cm} &
\hspace{0.1cm}  $\lambda^2$ ($\zeta_q$) \hspace{0.1cm}
\\ \hline
0.60  &  2.50 & 2.50 &  0.54 & 5.70 & 0.75 & 0.51 & 0.45 & 1.8 \\ \hline
0.65  &  2.85 & 2.85 &  0.79 & 5.27 & 0.70 & 0.74 & 0.57 & 2.45  \\ \hline
0.70  &  3.33 & 3.31 &  1.16 & 5.02 & 0.60 & 1.65 & 0.69 & 3.1 \\ \hline
0.75  &  4.0  & 3.92 &  1.73 & 4.72 & 0.54 & 3.26 & 0.85 & 3.7 \\ \hline
0.80  &  5.0  & 4.79 &  2.59 & 4.42 & 0.51 & 5.67 & 1.03 & 4.2 \\ \hline
0.85  &  6.66 & 6.05 &  3.95 & 4.03 & 0.50 & 6.83 & 1.24 & 5.5 \\ \hline
\end{tabular}
\end{center}
\caption[]{\small Value of different observables and fit parameters for different 
values of $z_2$, at fixed memory kernel ($\alpha=1.15$, or $\nu=0.3$). 
The values of $\lambda^2$ must be multiplied by $10^{-2}$.
Note the $10 \%$ discrepancy between the theoretical and 
empirical value of $\langle \sigma^2 \rangle$ when 
$z_2$ reaches $0.85$. The value of $\mu$ suggests that the kurtosis 
$\kappa$ remains finite at least up to $z_2=0.85$, 
in agreement with our theoretical analysis; the numerical value of 
$\kappa=3 {\cal F}(0)$ is found to be $\approx 12$ for $z_2=0.85$. From this table,
one can extrapolate $\mu$ to be $\approx 3.6$ and $\lambda^2$ ($\ln$) to be $\approx 0.015$
for $z_2=0.9$.
}
\end{table}

From Table I, we see that (a) the numerical value of the average volatility is close to 
its theoretical value
up to $z_2 \approx 0.75$, beyond which a systematic underestimation 
of the true volatility $\sigma^2$ is observed, which 
reaches $10 \%$ for $z_2=0.85$; (b) the kurtosis $\kappa=3 {\cal F}(0)$ increases with 
$z_2$, as expected, and seems 
to remain finite at least up to $z_2=0.85$, beyond which $\mu$ appears to drop below $4$, 
signaling a divergence of 
$\kappa$ (and correspondingly an even more difficult determination of the statistical 
properties of the system).

\subsubsection{Volatility correlations}

We now turn to the temporal correlations of the volatility. 
Several characterizations of the ``long-memory'' 
property are interesting to consider. Well studied quantities are correlations of 
different powers of the 
volatility, or of the logarithm of the volatility. 
In our model, we of course know exactly the volatility at
any instant of time, whereas, as pointed out above, in real conditions 
one only has access to price changes, from which a (noisy) proxy
of the volatility is constructed. 
We find numerically that the shape of the correlation function can be noticeably
different for these two quantities when the noise is large; this observation 
may be especially important for calibration. 

From Table I, one sees that the average volatility is rather ill-determined 
in the cases most relevant for applications, i.e. $z_2$ and $\alpha$
both close to unity, variograms should be preferred to correlograms \cite{Book}. 
In other words, we will study the following quantity:
\be
V_n(\ell) = \frac{1}{n^2}\langle \left(\sigma_i^n-\sigma_{i+\ell}^n\right)^2 \rangle.
\ee 
These variograms are plotted in Figs. 4 a,b for the case $n=2$ and $n \to 0$.
This last case reproduces, thanks to the $1/n^2$ normalization, 
the variogram of the logarithm of the volatility which
has been much studied in the context of multifractal models (see below). The case $n=2$ is 
important because it can be analytically studied, as we did in Section \ref{analytics}, 
and because it is related to the 
kurtosis of the distribution of returns for different time lags, 
which determines the smile of option prices. 
Form Eq. (\ref{full}), one finds that for large $\ell$, one should observe:
\be\label{theo_q2}
V_2(\ell) \sim 2{\cal F}_0 - 2 {\cal F}_\infty \ell^{-\nu} -2 {\cal F}_\infty'  \ell^{-2\nu} + ...
\ee
When $\alpha$ is close to 1, $\nu$ is small and one should a priori be prepared 
to see corrections to the 
asymptotic result coming from the $\ell^{-2\nu}$ contribution. 
Fig. 4-a however shows that, for $z_2=0.85$ and $\nu=0.3$ 
our numerical result is rather well fitted by the dominant 
term of Eq. (\ref{theo_q2}). The value of the 
apparent exponent $\nu$ however increases when $z_2$ decreases 
(in which case the contribution of the subleading term 
becomes more important). We also show the US stock data (that corresponds to 
$\nu \approx 0.25$ \cite{Book}), which can be matched quite well with the model.
In order to test the sensitivity of $V_2(\ell)$ to the value of $\alpha$, we also show in Fig. 4-a
the case $\alpha=1.3$, corresponding to $\nu=0.6$. The agreement with empirical data is clearly
not as good, a conclusion confirmed by all other observables we studied.

\begin{figure}
\begin{center}
\psfig{file=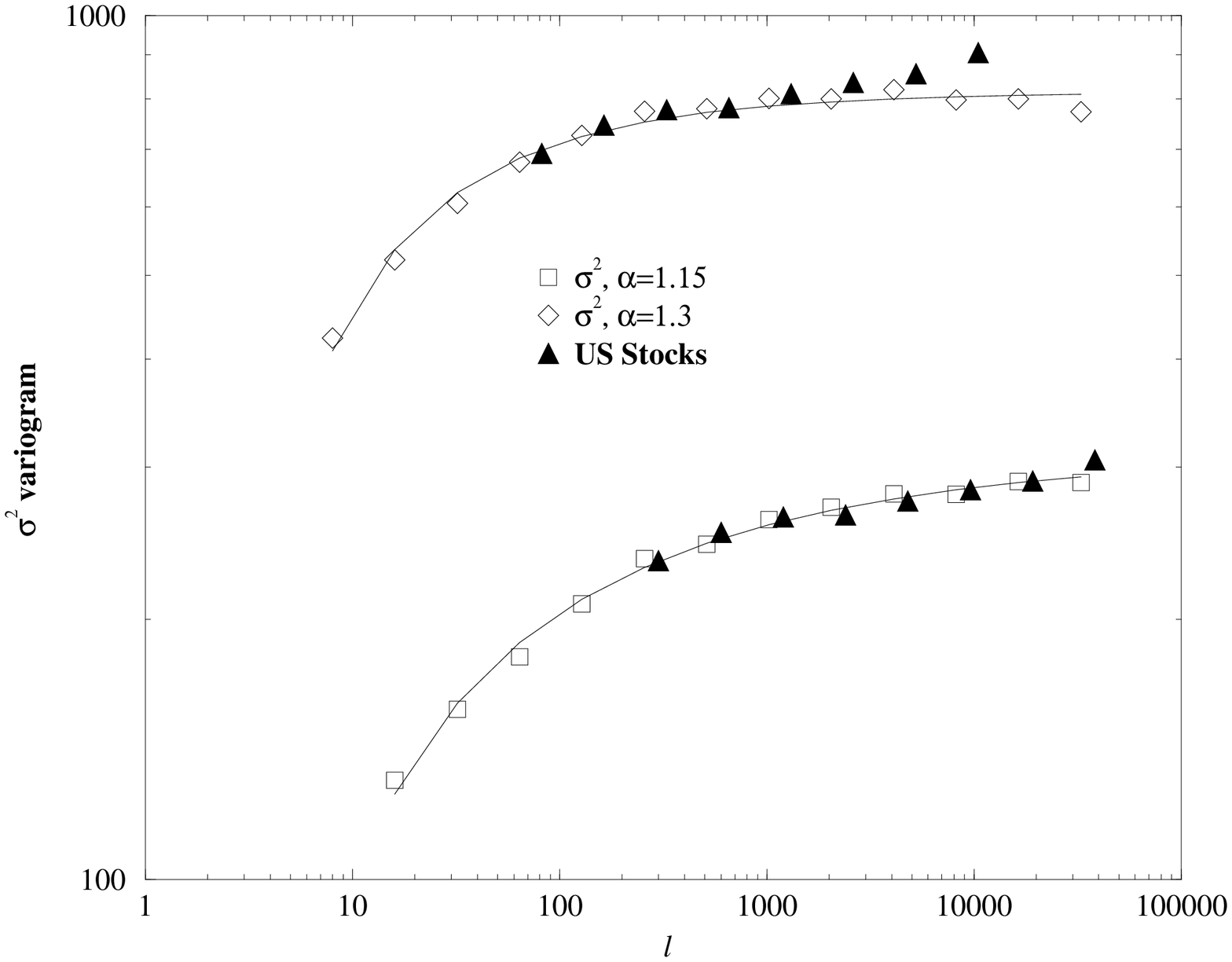,width=6cm,angle=270} \psfig{file=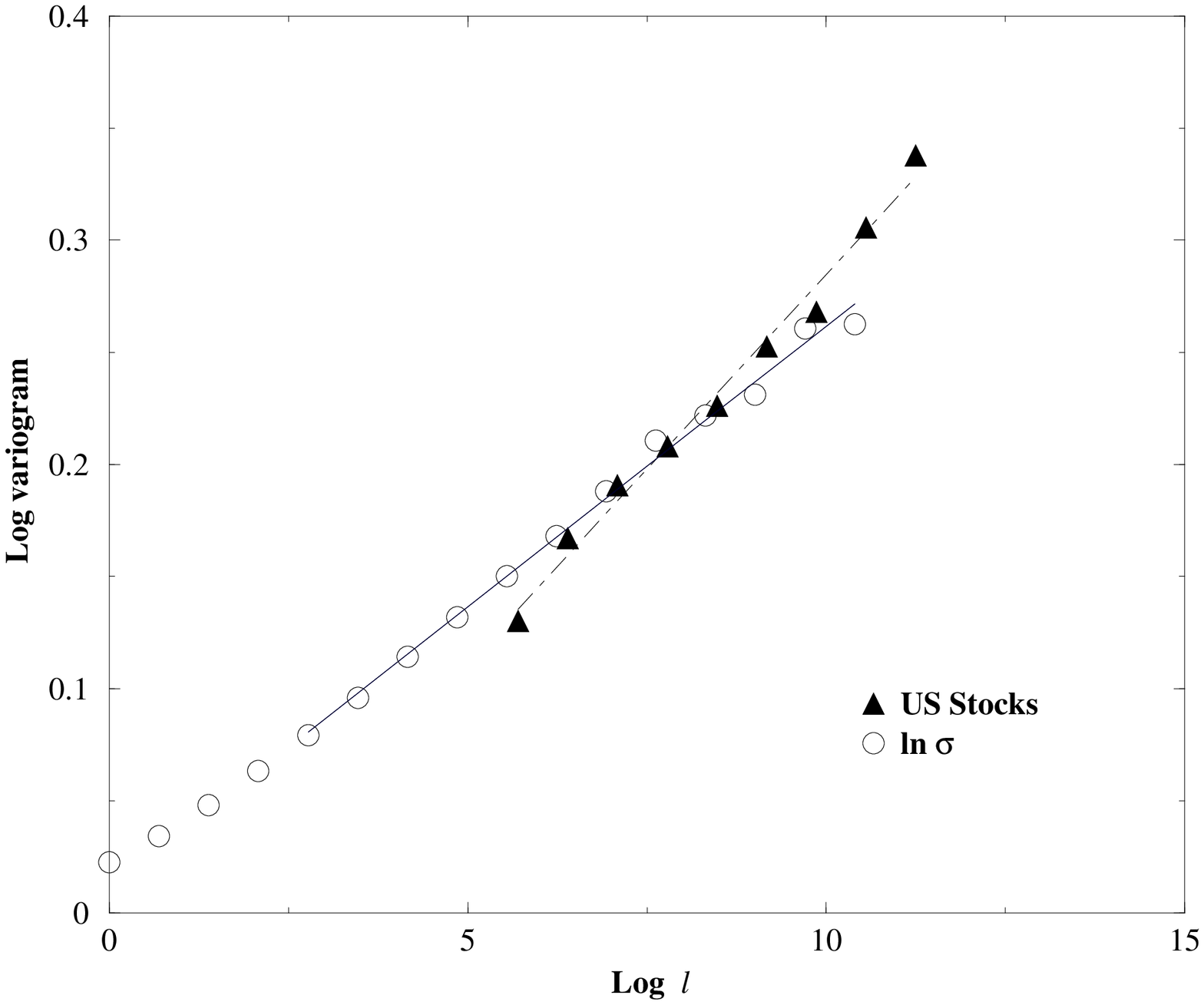,width=6cm,angle=270}
\end{center}
\caption{Left: variogram of the square volatility for $z_2=0.85$, 
$\alpha=1.15$ and $\alpha=1.3$ and fit with 
power-laws with $\nu=2(\alpha-1)$. We also show the data for US stocks with one day corresponding to 
$300 \tau$ for $\alpha=1.15$ and $80 \tau$ for $\alpha=1.3$, 
for which the agreement is clearly not as good. 
Right: variogram of the log-volatility for
$z_2=0.85$, $\alpha=1.15$, and fit with an affine function of $\ln \ell$, 
the slope of which yielding (twice) the 
intermittency parameter $\lambda^2$, here found to be $\approx 0.0125$, 
whereas the US data suggests a larger value $\lambda^2 \approx 0.0165$ -- 
which would be matched by choosing $z_2=0.90$, for which we estimate from Table I 
$\lambda^2 \approx 0.015$.}
\label{Fig4}
\end{figure}

Another interesting quantity, less noisy than the square volatility, 
corresponds to $n \to 0$. As discussed below, this
log-variogram appears naturally in the context of multifractal models. The result for $n=0$
is shown in Fig. 4-b; we see that it can be fitted approximately 
by the multifractal prediction \cite{BMD1}:
\be\label{theo_q0}
V_0(\ell) \sim 2 \lambda^2 \ln \frac{\ell}{L}, \quad (\ell \ll L)
\ee
where $\lambda^2$ is called the intermittency parameter, 
and $T=L\tau$ is usually called the integral time. 
The value of $\lambda^2$ for different values of $z_2$ is given in Table I, 
together with another determination of 
$\lambda^2$ discussed below. For $z_2=0.85$, $\alpha=1.15$, 
we find $\lambda^2 \approx 0.0125$, whereas our US data 
gives $\lambda^2 \approx 0.0165$, or 
$\lambda^2 \approx 0.018$ for the S\&P100 Index, given in \cite{MMS}.
This suggests that the optimal value of $z_2$ might in fact be closer to $0.9$,
for which we estimate from Table I 
$\lambda^2 \approx 0.015$. This conclusion is reinforced by the analysis of section VI.

\subsection{Distribution of returns over different time scales}

Since the noise variable $\xi$ is Gaussian, one can obtain 
the distribution of returns on the elementary
time scale $\ell=1$ from the distribution of the instantaneous volatility $\sigma$. For example,  
an inverse Gamma distribution for $\sigma$ leads to a Student-Tsallis distribution for $r$. 
As discussed above,
the actual distribution of volatility in our model appears 
to be slightly different from an inverse Gamma distribution;
therefore the distribution of returns in our model will be close to, but different from, 
a Student distribution. 
On larger time scales, the distribution progressively becomes Gaussian. 
However, the convergence is very slow
precisely because of the long-memory of the volatility, 
parameterized by the exponent $\nu$. A way to quantify this
convergence is to measure the cumulants of the distribution, 
for example the excess kurtosis $\kappa(\ell)$, expected
from our theoretical analysis to decay as $\ell^{-\nu}$, 
or the rescaled mean absolute deviation $\Upsilon(\ell)$, 
defined as:
\be
\Upsilon(\ell) = \frac{1}{\sqrt{\langle \sigma^2 \rangle \ell}} \langle |x_{i+\ell} - x_i| \rangle.
\ee
For a Gaussian distribution, one should find $\Upsilon=\sqrt{2/\pi}$. 
These quantities are plotted as a function of
$\ell$ in Figs. 5-a,b, for different values of $z_2$ and for $\nu=0.3$. 
An a priori unexpected feature is 
that non-Gaussian
effects actually first {\it increase} for small $\ell$, before decaying back 
to zero beyond a certain $\ell=\ell^* \approx 50$. 
The origin of this non monotonicity was discussed in section \ref{analytics} 
and is clearly related to the assumption that
the noise $\xi_i$ is Gaussian. Any extra kurtosis coming from unpredictable 
jumps in the price, not captured by
the feedback mechanism of our model, will strongly affect the shape 
of $\Upsilon(\ell)$ and $\kappa(\ell)$ on 
short time scales, and remove this non-monotonicity which, to the best of our knowledge, 
is not observed on 
empirical data, even on very short time scales. 
Another possibility is to change the shape of $g_\ell$ for small $\ell$'s.

\begin{figure}
\begin{center}
\psfig{file=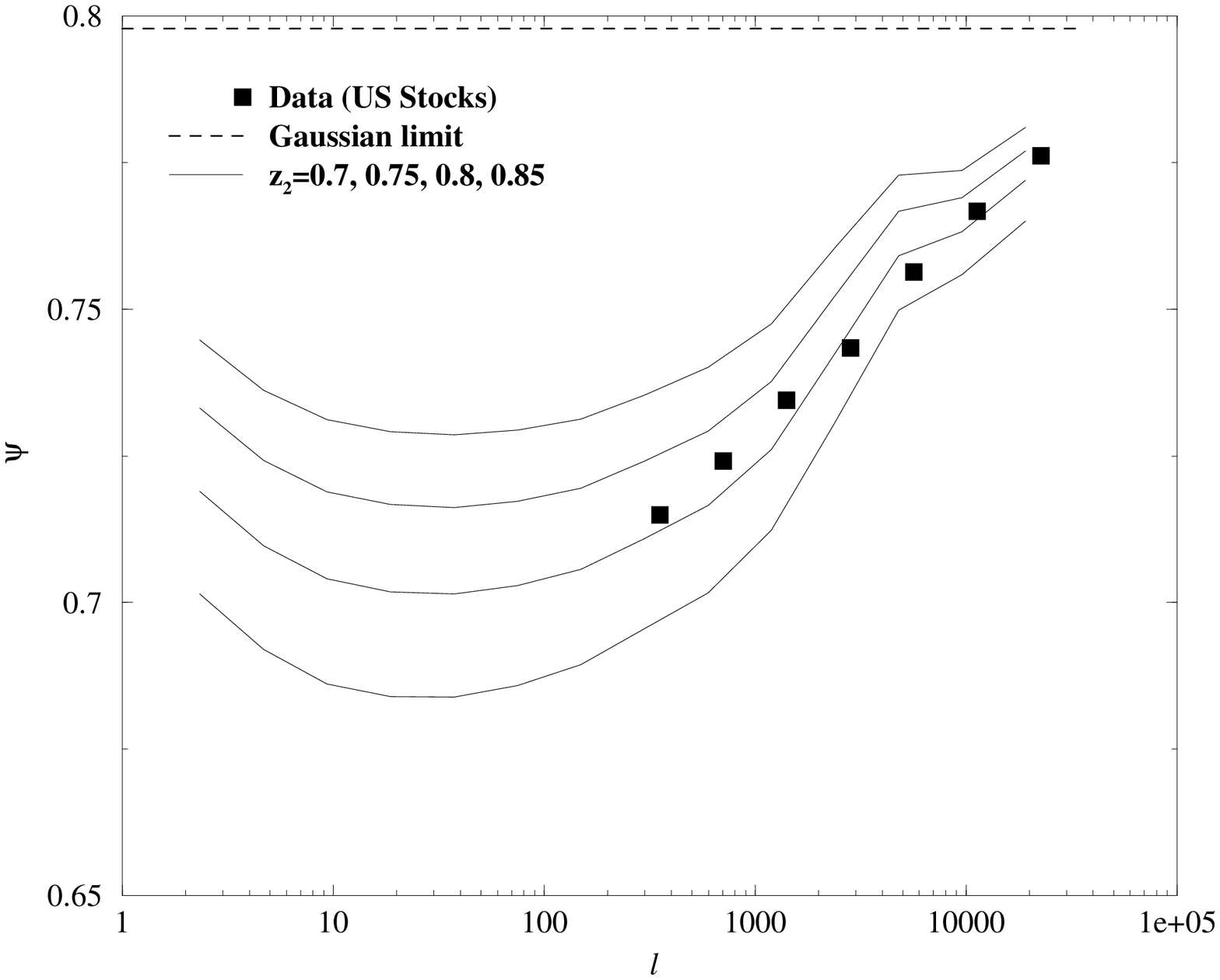,width=6cm,angle=270} \psfig{file=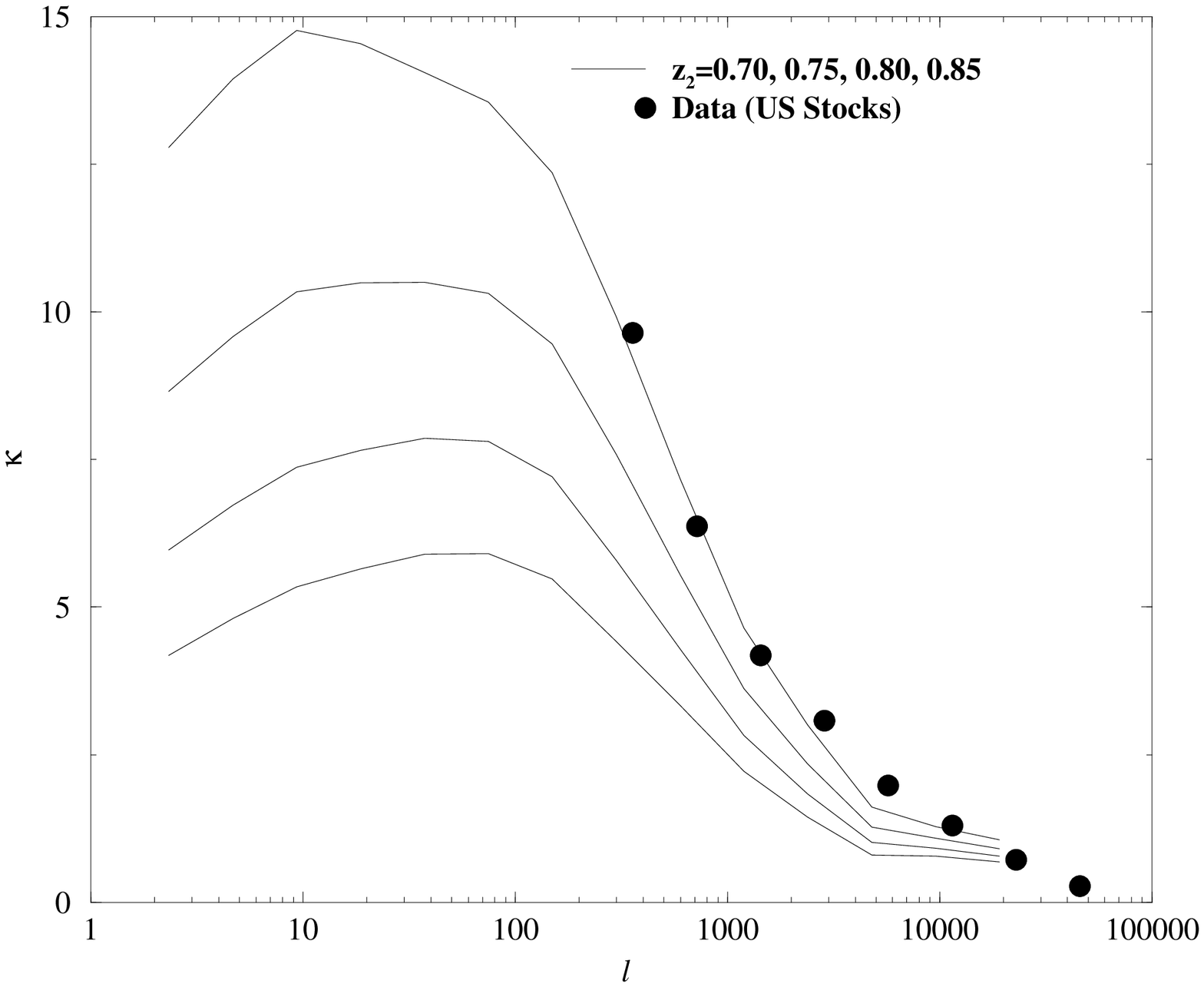,width=6cm,angle=270}
\end{center}
\caption{Evolution of two cumulants of the distribution of returns with the lag $\ell$, 
for different values 
of $z_2$. Left: rescaled mean absolute deviation $\Upsilon(\ell)$. Note that 
the evolution is non-monotonous as a function of $\ell$. 
Right: Excess kurtosis $\kappa(\ell)$. We have
also shown (symbols) the corresponding cumulants for US stocks, 
where we choose $\ell=300$ to correspond to one trading day, as in Fig. 4.}
\label{Fig5}
\end{figure}

Of course, the knowledge of $\kappa$ and $\Upsilon$ is insufficient to fully 
characterize the whole distribution on 
different time scales. We have in fact found that a 
Student-Tsallis distribution with a time dependent number of
degrees of freedom is an acceptable fit of this distribution for all values of $\ell$. 
In line the notation of 
ref. \cite{LB1}, we write this distribution as:
\be\label{student}
P_\ell(\Delta) = {\cal N} \frac{\Delta_0^{(3-q)/(q-1)}}{(\Delta_0^2 + 
(q-1) \Delta^2)^{1/(q-1)}},\qquad \Delta=x_{i+\ell}-x_i
\ee
with an $\ell$ dependent parameter $q(\ell)$. In the limit $q \to 1$, the distribution becomes Gaussian. 
If the 
distribution is indeed given by Eq. (\ref{student}), the relation between $\Upsilon$ and $q$ reads:
\be
\Upsilon=\sqrt{\frac{2 (\mu-2)}{\pi (\mu-1)^2}} \frac{\Gamma(\frac{1+\mu}{2})}
{\Gamma(\frac{\mu}{2})}, \qquad \kappa=\frac{6}{\mu - 4},
\ee
with $\mu \equiv (3-q)/(q-1)$. Using these relations, one can infer, 
from Fig. 5-a, the value of $q$ that one should use for different times scales
in order to get an {\it approximate} functional form for the distribution of returns. 
This is useful for option pricing, for example \cite{LB1}.

\subsection{Multifractality}

A property related to the systematic change of the distribution of returns with 
$\ell$ is {\it multifractality}, which
means that different moments of price changes scale as a power of time, but with different scaling 
exponents. More precisely, multifractal scaling is the following property:
\be\label{moments}
M_n(\ell)=\langle |x_{i+\ell} - x_i|^n \rangle = A_n \ell^{\zeta_n}; \qquad \ell \ll L
\ee
where $A_n$ are constants and $\zeta_n$ is an $n$-dependent exponent. In the monofractal case, where the 
distribution is the same on all time scales up to a rescaling of the returns,
then $\zeta_n = n/2 \, \zeta_2$. 
The simplest example is obviously the (geometric) Brownian motion, for which 
$\zeta_n= n/2$. Any deviation from a linear behaviour of $\zeta_n$ is coined multifractality, for which 
several explicit models were proposed recently \cite{Lux1,Lux2,CF1,CF2,BMD1,BMD2}.
 
One example is the
Bacry-Muzy-Delour ({\sc bmd}) stochastic volatility model, 
which makes the following assumptions \cite{BMD1,BMD2}:

\begin{itemize}
\item the log-volatilities $\ln \sigma_i$ are multivariate Gaussian variables (or more generally infinitely
divisible \cite{BMD2}).

\item the log-volatility variogram is given by Eq. (\ref{theo_q0})

\item the volatilities $\sigma_i$ are independent from the (Gaussian) noises $\xi_i$.
\end{itemize}

From these assumptions, one can compute exactly the moments of the return distribution on different time scales. One
finds that these are indeed given by Eq. (\ref{moments}), with:
\be \label{zetan}
\zeta_n=\frac{n}{2} [1-\lambda^2(n-2)],
\ee
whenever $n < 1/\lambda^2$, beyond which the moments are infinite. (All $A_n$'s can also be exactly computed \cite{BMD1}). 
These assumptions and predictions were found to account rather well for some aspects of empirical data.

In the present section, we show that although our model is, strictly speaking, {\it not} multifractal,
many of the multifractal predictions actually hold numerically quite accurately. 
This means that our model can
account very well for apparent multifractal properties of financial time series, and in fact cures some 
of the deficiencies of standard multifractal models (see below). First, we note that our model is not 
multifractal since 
the moments $M_n(\ell)$ can be exactly computed to be {\it sums} of power-laws with different
exponents, and not a unique power-law (see \cite{BPM} for a related discussion). For example, $M_4(\ell)$ is 
the sum of $\ell^2$, $\ell^{2-\nu}$, $\ell^{2-2\nu}$, etc., and therefore 
does not scale as a unique power-law. 
However, as we show now, the numerical behaviour appears 
difficult to distinguish from a unique, effective power-law.
\footnote{The marginal case $\alpha=1$ with a cut-off for $\ell > L$ 
is quite interesting and might in fact resemble 
the strictly multifractal {\sc bmd} model.} 

\begin{figure}

\begin{center}
\psfig{file=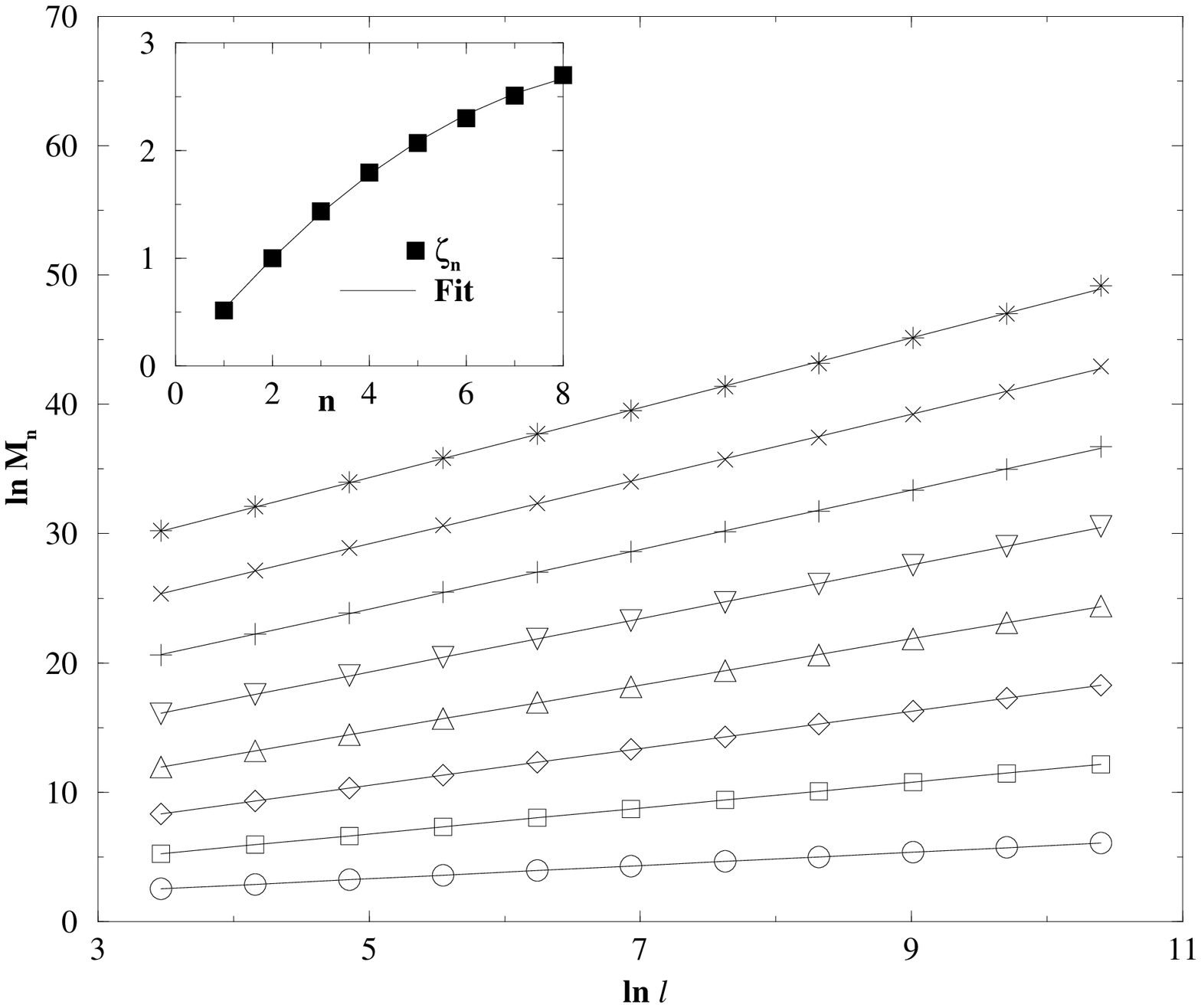,width=5cm,angle=270} \psfig{file=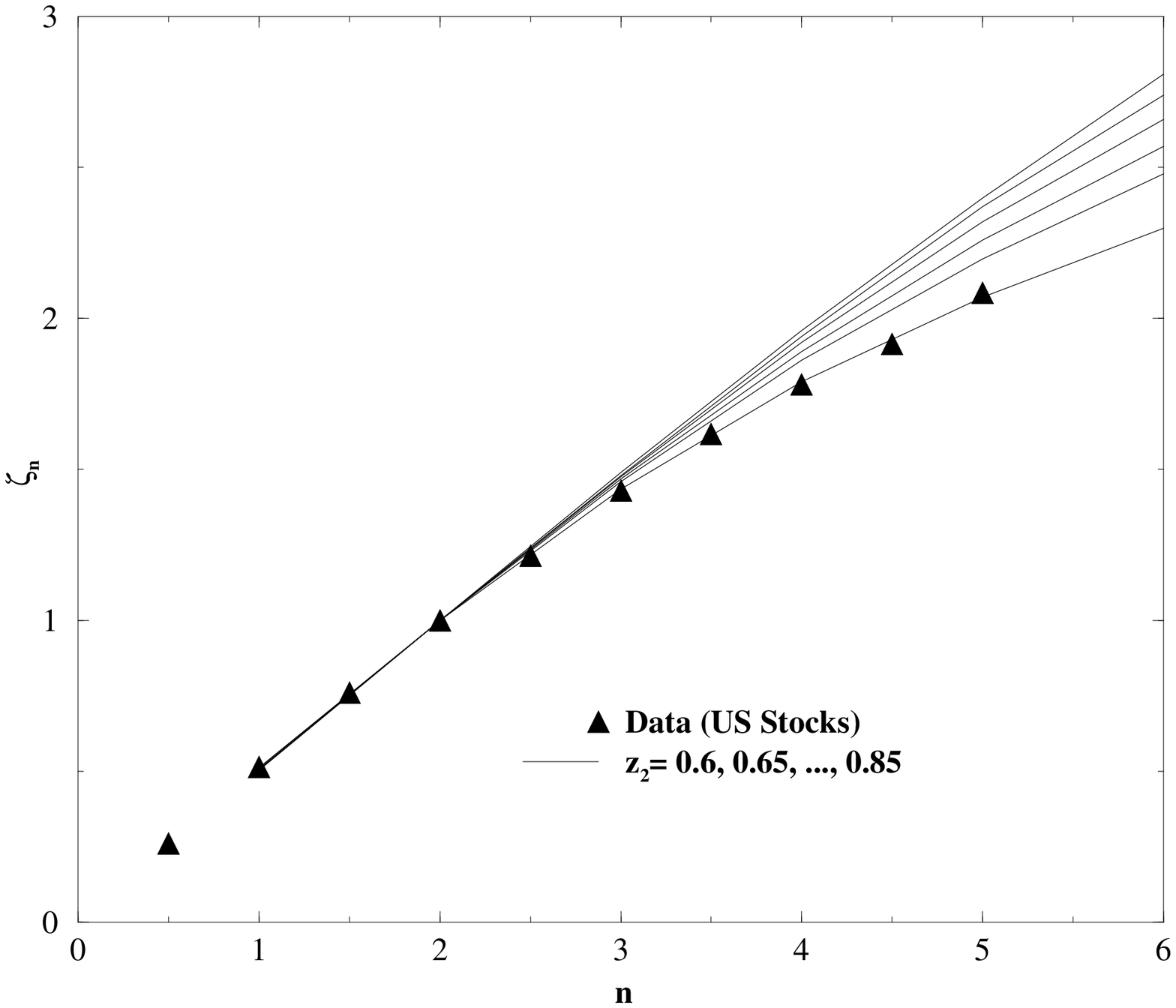,width=5cm,angle=270}
\end{center}
\caption{Left: Evolution of different moments $M_n(\ell)$ of our model with $\ell$ in a log-log representation, which allows one 
to extract from the slope of these lines, the exponents $\zeta_n$ shown in the inset. Also shown in the inset is
the parabolic fit suggested by the {\sc bmd} model, Eq. (\protect\ref{zetan}). Right: The exponents $\zeta_n$ as
a function of $n$ for different values of the feedback parameter $z_2$, and the corresponding results of US stocks 
(triangles). 
}
\label{Fig6}
\end{figure}

We have computed numerically $M_n(\ell)$ for $\ell > \ell^*$, where $\ell^*$ corresponds to the maximum of 
$\kappa(\ell)$ or
the minimum of $\Upsilon(\ell)$ appearing in Figs. 5-a,b. In this regime, one can neglect the contribution of terms
involving $\langle \sigma_i^2 \xi_j^2 \rangle_c$, and $M_n(\ell)$ can be expressed as:
\be
M_n(\ell) \approx (n-\frac12)!! \langle \left(\sum_{i=1}^\ell \sigma_i^2\right)^{n/2} \rangle,
\ee
which is the quantity that we studied numerically, because it is much less noisy than the direct calculation
of moments of returns. The results are shown in Fig. 6-a in a log-log representation, for $z_2=0.85$ and $\alpha=1.15$,
from which it is obvious that pure power-laws are indeed excellent fits. From the slope of these lines one obtains
the exponents $\zeta_n$, shown for different values of $z_2$ in Fig 6-b. We note that:

\begin{itemize}

\item A parabolic fit of $\zeta_n$ as a function of $n$, as in Eq. (\ref{zetan}), 
gives an excellent representation
of our data (see Fig 6-a, inset).

\item From this parabolic fit, a value of $\lambda^2$ can be extracted for different values of $z_2$. This 
value of $\lambda^2$ is four times larger than that extracted from the variogram of the log-volatility, in
contradiction with the {\sc bmd} model, where both should be equal (see Table I). However, we note 
that a similar discrepancy with the {\sc bmd} model is observed on US stock data as well, but that 
both observables are fully compatible with our model with the same set of parameters. 
The discrepancy with the log-normal {\sc bmd} model is due to the 
underestimation of the probability of large events in that model.

\item The intermittency parameter $\lambda^2$, that 
gauges the degree of multifractality (i.e. the deviation of
$\zeta_n$ from a straight line), increases as $z_2$ increases. 
The multifractal spectrum extracted from US stock data 
corresponds to $\lambda^2 \approx 0.055$ and matches quite well our 
numerical points for $z_2=0.85$. Similar values of
$\lambda^2$ have been reported for other markets as well (see e.g. \cite{multif2,BMD1}).

\end{itemize}

The {\sc bmd} multifractal model makes other, even more detailed predictions, 
about the relaxation of volatility 
after a volatility shock. We now turn to this topic to show that our model can also reproduce these more 
subtle features. 

\subsection{Response to volatility shocks}

A question of great importance for option pricing and risk management concerns `aftershocks'. 
It is well known that 
after a large market move, the volatility remains high for a while. 
The precise question therefore is: conditioned to
a large volatility burst, how fast will the market revert back to normal? 
This has been addressed both empirically and
theoretically, within the context of the {\sc bmd} model \cite{MMS}. 
One finds that after a shock, the volatility reverts 
to its normal level very slowly, as a power-law of the time $\ell$ after the shock:
\be
\Delta \sigma_{i+\ell} \sim \Delta_0 \ell^{-\theta},
\ee
where $i$ is the time of the initial shock, $\Delta \sigma$ the excess volatility over its average value, 
and $\Delta_0$ the amplitude of the initial shock. For rather large shocks, the empirical data suggests 
$\theta \approx 1/2$ \cite{MMS,kertecz2}, while $\theta$ decreases for smaller shocks. 
Interestingly, the multifractal 
{\sc bmd} model suggests that the exponent $\theta$ in fact depends continuously on the amplitude of the 
initial shock, and decreases from the value $1/2$ as the amplitude of the shock decreases \cite{MMS}. 
This prediction 
was found to be in remarkable agreement with empirical findings, 
giving strong support to the {\sc bmd} picture. 

\begin{figure}
\begin{center}
\psfig{file=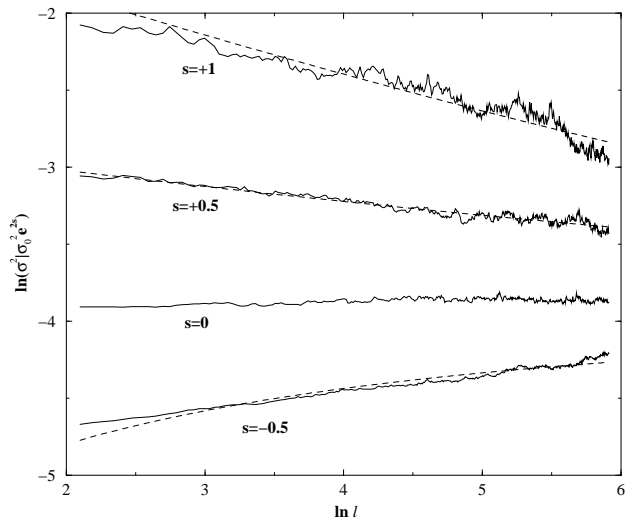,width=7cm,angle=270}
\end{center}
\caption{Evolution of $\sigma^2$ conditioned to an initial volatility 
$\langle \sigma^2 \rangle e^{2s}$, with $s=-1/2,0,1/2,1$
(from bottom to top). The dashed lines are power-law fits, 
with exponents $\theta(s=-1/2) \approx 0.22$, $\theta(s=1/2) \approx 0.17$
and $\theta(s=1) \approx 0.30$, very similar to the results quoted in \cite{MMS} for the S\&P100.}
\label{Fig7}
\end{figure}

We have therefore computed the volatility relaxation process within our model, 
following the methodology of 
\cite{MMS}. We compute the average volatility a time $\ell$ after the shock, 
conditioned to a shock of a certain amplitude. 
The relaxation curves are shown in Fig 7, again in the case $z_2=0.85$, $\alpha=1.15$. 
We observe that the predictions 
of the {\sc bmd} model are again quite accurately verified by our model, which, by the same token, 
is an alternative candidate to explain empirical results. 

Another characterization of aftershocks inspired from research on earthquakes is the 
so-called Omori law, which
states that the probability of an aftershock larger than a certain threshold 
occurring a time $\ell$ after the 
main shock decays as $1/\ell^p$, with $p \approx 1$. 
This law was checked for stock markets in \cite{MantegnaOmori}
on a handful of `significant' crashes. In our model, crashes are self-generated 
and not related to external news, 
obviously absent from the model. We show in Fig. 8 the numerically
determined Omori law after large endogenous crashes, 
for which we obtain a significantly different value of $p \approx 0.5$,
compatible with the value of $\theta$ reported above for large crashes. 
On the other hand, it is easy to compute the
volatility response to an exogenous crash, represented by a large instantaneous 
jump added `by hand' in the time series. If the amplitude of the jump at time $i=0$ is $J$, 
the volatility after the crash is given by:
\be
\langle \sigma_\ell^2 \rangle_J \approx \langle \sigma^2 \rangle + g J^2 M_\ell.
\ee
Using $M_\ell \sim \ell^{-\alpha}$, we find that the probability of an aftershock larger 
than a given threshold also decays as 
$\ell^{-\alpha}$ for large enough $\ell$. 
Since the value of $\alpha$ is close to unity, an approximate Omori law with $p \approx 1$ 
will be observed after 
anomalously large, exogenous crashes in our model. 
The distinction between endogenous and exogenous crashes, suggested
in \cite{MMS}, makes perfect sense in the context of the present model, 
where endogenous crashes are, in a precise sense,
the result of progressive volatility built up, resulting from the ARCH like feedback effect. 
This volatility built up is
in fact related to the non monotonous behaviour of the kurtosis in our model.

\begin{figure}
\begin{center}
\psfig{file=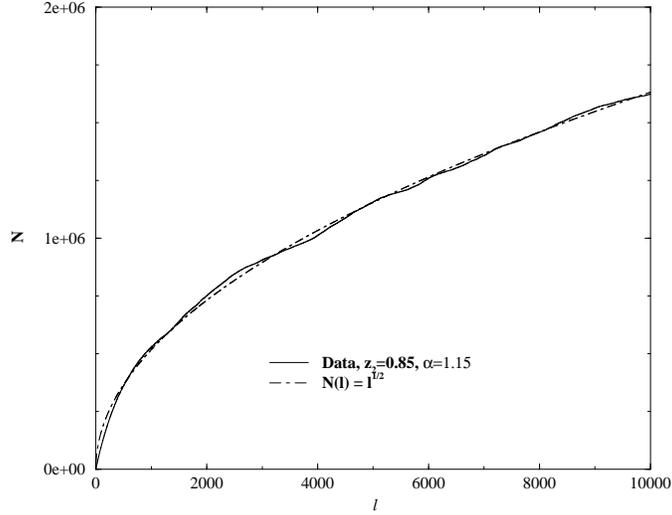,width=7cm,angle=270} 
\end{center}
\caption{This Omori plot shows the cumulative number of aftershocks 
(i.e. returns with an amplitude larger than a certain
threshold) following a main shock, and fit with $N(\ell) \sim \sqrt{\ell}$. 
Main shocks were defined as returns larger than a third of the maximum return 
observed over the whole time series (of length 850,000), 
and aftershocks as returns larger than a third of the main shock.}
\label{Fig8}
\end{figure}

\subsection{Time reversal symmetry}

A question of general interest is whether financial time series `know' about the arrow of time, 
i.e. whether it is
possible to compute any observable that distinguishes past from future (see \cite{Pomeau}). 
Although the answer of 
this question would 
appear, to the layman, to be trivially yes, things turn out to be much more subtle, 
and of considerable 
importance. For example, the usual Brownian 
motion, all L\'evy processes and all multifractal models constructed up to now 
(including Mandelbrot's cascade, the {\sc bmd} model or the version studied 
by Lux in \cite{Lux2}) are strictly invariant under time reversal symmetry ({\sc trs})! 
Financial data, on the other hand, do 
reveal non {\sc trs} effects. A simple example, on which we will expand 
in the next section \ref{leverage}, is the
leverage effect, which is a causal correlation between 
{\it past} price changes and {\it future} volatilities: 
a drop in price leads to an increased volatility. 
This effect in turn leads to some (negative) skewness in the distribution
of returns (see below). 

Here, we want to discuss a distinct effect, recently evidenced by Zumbach and Lynch \cite{LZ}.
In order not to mix this effect with leverage, one can study FX rates between two large currencies, 
for example 
Euro vs. Dollar. 
In this case, any leverage correlation or skewness, if present, is very small. 
In spite of this, there is a clear time asymmetry in the volatility process: 
as shown in \cite{LZ}, the correlation between large scale, past 
volatilities and small scale future volatilities is larger than between small scale, past 
volatilities and large scale future volatilities. This effect was also noted in \cite{Arn}, 
but on the example of the 
S\&P 500 index for which the leverage effect is very strong. 
We have computed this correlation in our model, following 
the methodology outlined in \cite{LZ}, where the idea of `mug-shots' was introduced to
represent graphically such past volatilities/future volatilities correlations. The 
mug-shot corresponding to our model is shown in Fig. 9.
It is clear that
our model -- almost by construction -- captures such a non {\sc trs} effect. 
This was already noted in \cite{LZ} for a similar model.

We think that the time asymmetry revealed by Zumbach's mug-shots is extremely important: 
first, it imposes a
theoretical constraint on the eligible models of financial time series that most of them fail to obey. 
Second, it
is a direct proof of the existence of feedback effects in financial markets: 
the history of past price changes 
does have a direct impact on the decision and behaviour of traders -- in plain contradiction with the  
efficient markets dogma.

\begin{figure}
\begin{center}
\psfig{file=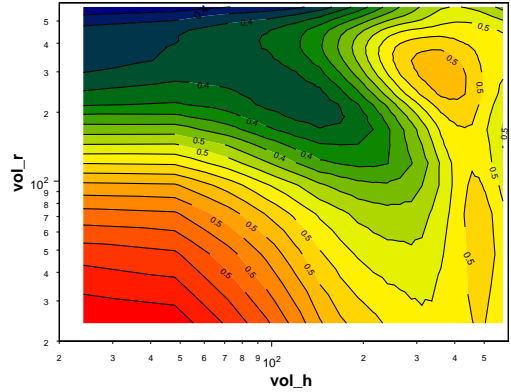,width=8cm} 
\end{center}
\caption{Zumbach's mugshot for our model: contour plot of the correlation between past volatility and future 
volatility, measured on different time scales. For a {\sc trs} process, this mug-shot would appear 
symmetric around the diagonal, whereas empirical data shows, as in this figure, that the region 
below the diagonal carries more correlation than the region above it.}
\label{Fig9}
\end{figure}

\section{The leverage effect}
\label{leverage}

Up to now, we have 
only discussed our feedback model under the assumption of a symmetric reaction of the market 
participants to price changes. If negative price changes have a larger impact than positive price changes, 
i.e,
if the distribution of thresholds shown in Fig. 1 has some asymmetry, 
one will observe negative correlations 
between past price changes and future volatilities (leverage effect) 
and some skewness in the distribution of returns,
totally absent from the above model. 
The natural way to generalize Eq. (\ref{the-model}) to account for such an
asymmetry is to write:
\be\label{model-leverage}
\sigma_i^2 = \sigma_0^2 \left[1 + \varphi \sum_{\ell=1}^\infty 
g_\ell \frac{(x_i-x_{i-\ell})}{\sigma_0 \sqrt{\ell \tau}}
+\sum_{\ell=1}^\infty 
g_\ell \frac{(x_i-x_{i-\ell})^2}{\sigma_0^2 \ell \tau}\right],
\ee
where $\varphi$ measures the strength of the asymmetry. 
The case $\varphi=0$ reproduces the model studied above,
while $\varphi < 0$ induces a leverage effect. 
A sufficient condition on $\varphi$ that ensures that $\sigma^2$ always remains 
positive is to impose that each term of the sum over $\ell$ contributes positively. 
Writing as an identity $1 = \sum_{\ell=1}^\infty g_\ell/z_2$, one obtains:
\be
X^2 + \varphi X + \frac{1}{z_2} \geq 0 \qquad \forall X,
\ee
or: $z_2 \varphi^2 < 4$.

The leverage correlation can be defined as:\cite{BMaP}
\footnote{A slightly different normalization was actually used in \cite{BMaP}.}
\be
{\cal L}(i-j)=\frac{1}{\langle r^2 \rangle^{3/2}} \langle r_i^2 r_j \rangle.
\ee
This quantity is found empirically to be close to zero for 
$i \leq j$ and negative for $i > j$. It is not difficult to compute exactly 
this correlation function in our model, provided the distribution of $\xi_i$ is even. 

One finds:
\be
\langle r_i^2 r_{i-\ell} \rangle = 
\langle \sigma^2 \rangle \sigma_0 \tau^{3/2} \varphi \sum_{j=\ell}^\infty 
\frac{g_j}{\sqrt{j}}.
\ee
One should also note that the average volatility is unchanged by the leverage term $\varphi$. 
Therefore, using $\langle \sigma^2 \rangle=\sigma_0^2/(1-z_2)$, we finally find:
\be
{\cal L}(\ell) = \varphi g \sqrt{1-z_2} \sum_{j=\ell}^\infty
\frac{1}{j^{1/2+\alpha}} \sim_{\ell \to \infty} \ell^{1/2 - \alpha}.
\ee
The decay of the empirical leverage correlation with lag,
although noisy, can be fitted by a power-law of exponent close to $0.5$,
not far from $\alpha-1/2$ (see Fig. 10).
A power-law decay of the leverage correlation was also proposed in the context of the multifractal 
{\sc bmd} model in \cite{Pochart,kertecz}.

\begin{figure}
\begin{center}
\psfig{file=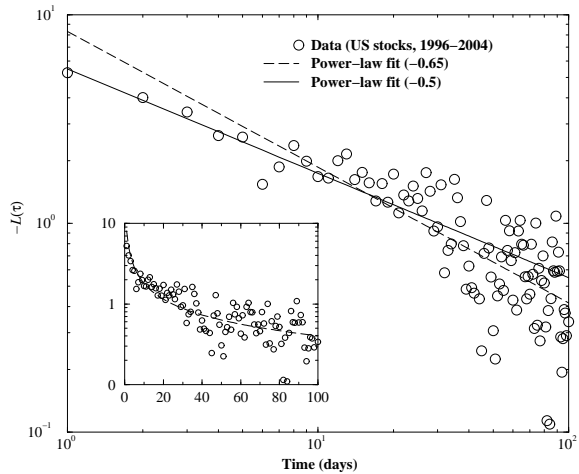,width=7cm,angle=270}
\end{center}
\caption{
The leverage correlation ${\cal L}$ (with a minus sign) as a function of lag, for US stocks, in a 
log-log representation. The straight line corresponds to the best power-law fit over the whole 
range and has slope $-1/2$, whereas the prediction of our model for $\alpha=1.15$ is a slope of $-0.65$ 
(dashed line). Note however the scatter in the empirical data points: the leverage effect for stocks is weak 
and hard to measure \cite{BMaP}. Inset: same data in a linear representation, with the prediction of our
model.
}
\label{Fig10}
\end{figure}

This quantity is important because it governs the behaviour of the skewness 
of the return distribution on different time scales. 
It is indeed easy to show that the normalized skewness of returns on scale $\ell$, 
${\cal S}(\ell)$ is given by \cite{Book}:
\be
{\cal S}(\ell) = \frac{3}{\sqrt{\ell}} \sum_{j=1}^{\ell} (1 - \frac{j}{\ell}) 
{\cal L}(j) \sim_{\ell \to \infty} \ell^{1 - \alpha} \qquad (1 < \alpha < 3/2).
\ee
From the above expression, we see that even if the return distribution is symmetric on the smallest 
time scale (${\cal S}(1)=0$), a negative skewness appears for $\ell > 1$ when $\varphi < 0$, 
and decays back to zero for very large lags. 
However, once again, the proximity of the critical line $\alpha=1$ beyond which 
the process becomes non stationary, leads to a very slow decay of the skewness. 
Empirically, for daily returns of 
individual stocks, one finds ${\cal S} \approx -0.1$, corresponding to $\varphi \sim - 1$ 
when $\alpha=1.15$ and $z_2=0.85$.

The skewness of stock indices, on the other hand, is generally much larger (by a factor $10$) 
than that of individual stocks. This is due 
to an enhanced downside correlation, which should be modeled using the multi-asset model discussed below.

Note that the extra asymmetric term introduced in this section actually 
contributes also to the volatility-volatility
correlation ${\cal F}$ computed above and also to the kurtosis. 
For large $\ell$, this extra contribution behaves as 
$(1-z_2)\varphi^2/\ell^\nu$ and adds to the dominant term computed in section \ref{analytics}.

\section{Soft calibration with real time series}
\label{calibration}

We have shown in the above sections, using both analytical arguments and numerical simulations, that 
our model Eq. (\ref{the-model}) is able to reproduce semi-quantitatively many of the stylized facts of 
financial time series that have been reported and studied in the literature. We have in fact shown, in many
of the above figures, empirical data that match quite well, at least to the eye, 
the predictions of the model. What do we mean by `semi-quantitatively'? 
Can one be more quantitative and calibrate, in a standard econometric sense, our model to empirical data? 

We believe that our model is interesting precisely because it clearly underlines 
the limits of such an ambition. 
The empirical data clearly suggests that any faithful statistical model of financial time series 
must be somehow close to being non-stationary.
This is obvious from the very existence of option markets, which demonstrate the difficulty of measuring and
predicting the volatility, even on rather long periods: the at-the-money vol of long-dated options still 
moves around
quite a bit from day to day and there is a persistent smile, symptomatic of a long-memory extending to a 
few years \cite{Book}. We have shown in the above
figures that empirical data on stocks seem to favor values of $z_2$ and $\alpha$ that drive our model 
very close to
instability. This means that even a million step long simulation of our theoretical model for 
realistic parameters is
insufficient to determine the true value of the volatility to better than $5\%$ (see Table I); 
such an uncertainty 
affects all the observables of the model. How can one believe that anything
more precise than this can be reached on real empirical data? 
Available time spans are necessarily restricted, 
true jumps and overnight effects make the returns even more kurtic, true seasonalities (day, week, 
month, quarters, years) certainly play a role, and non-stationarities (for example, the acceleration of the
trading frequency with time) plague any attempt to represent the dynamics of financial markets with fixed 
values 
of the parameters on very long time scales. No test, and no model, should aim at more precision than 
reality itself.

In this situation, we think that the only reasonable strategy is what one could call `soft calibration',
in the following sense:
instead of focusing on a few observables that one tries to reproduce as accurately as possible to calibrate 
the model (which is always possible), one should instead find a set of parameters that 
approximately accounts for {\it as many} different observations 
as possible, and cross check the overall consistency of the model. This consistency is more 
important and more stringent than a perfect 
fit of an intrinsically elusive target. Calibration in these extreme conditions is an ill-posed problem 
that, we believe, must be supplemented by intuition on what is important and plausibility. 

\subsection{Calibration on stylized facts}

How does this work in practice for the model we studied? In the simplest version that we developed, the model has 
three important parameters: $\tau$, $z_2$, $\alpha$ (the value of $\sigma_0$ merely sets the scale of the returns, 
but has no bearing on the structural properties of the model). 
Accounting for the leverage effect adds one more parameter,
$\varphi$. But we already know, both from the numerical results that show that our model 
leads to a non-monotonous 
kurtosis, and from common sense, that unpredictable jumps must be present and should be factored in through 
a non-Gaussian noise term $\xi$, which most probably has itself fat, power-law tails \cite{LZ,Farmer}. 
This adds at least another parameter, which would play an important role in an extended formulation of the model. 
Neglecting for now this extra complication, our strategy is based on the idea that
different observables probe differently the influence of all parameters. 
This is if fact how we organized the numerical results of section \ref{numerics}:

\begin{itemize}

\item The distribution of the volatility or of the returns probes primarily the value of $z_2$. The tail exponent $\mu$,
and any measure of non-Gaussianity helps restricting the range of acceptable values of $z_2$ (see Fig. 3 and Table I).

\item The temporal correlations of the volatility is primarily sensitive to the value of $\alpha$, 
and can be used to
limit the acceptable range of this parameter (see Fig. 4-a), whereas the correlation of the log-volatility is most
sensitive to the value of $z_2$ (see Fig. 4-b and Table 1).

\item The evolution of the non-Gaussian cumulants $\kappa$ and $\Upsilon$ is sensitive 
to $z_2$, $\alpha$, but also 
to the value of the elementary time scale $\tau$ (see Figs. 5-a,b). 
This has enabled us to fix $\tau \approx 1/300$ day 
to be consistent with empirical data. Of course this leaves us 
with the task of curing the unfriendly looking 
short scale kurtosis, but as mentioned above, this 
could be easily be dealt with a non-Gaussian noise $\xi$. This however 
means that the optimal value of $z_2$ would be slightly reduced, 
since part of the kurtosis would already be accounted for.

\item The multifractal analysis provides a stringent 
cross-check of the choice of parameters, since the multifractal
spectrum $\zeta_n$ is quite sensitive to the value of $z_2$ (see Figs 6-a,b).

\item The consistency of the model can be probed further by analyzing 
the response of the volatility to shocks of
different amplitudes, and studying the Omori plots (see Figs 7, 8). 
An acceptable description of this rather 
subtle statistics is, we believe, another useful constraint on the parameter range.

\item Interestingly, the leverage correlation is totally decoupled from other observables and 
can be determined 
independently from the study of the asymmetry of the distribution of returns 
and asymmetric volatility correlations, 
that allow one to fix the parameter $\varphi$. 
[Note however that $\varphi \neq 0$ adds a contribution to the kurtosis of the process].

\end{itemize}

Following these steps is how we `calibrated' our model on the average behaviour of 
252 liquid US stocks in the four-year 
period 2000-2003. From Figs. 4-6, we see that the value $\alpha = 1.15$ allows one 
to capture correct time 
dependence of the volatility correlation and of the evolution of non-Gaussian cumulants, whereas the choice of $z_2$
in the range $0.85 - 0.90$ allows one to capture the correct level of non-Gaussianity and multifractality (the parameter
$\lambda^2$ appearing in Table I and in Figs 6). These values of $z_2$ and $\alpha$ allow us to reproduce quite
satisfactorily the whole set of observables that we have studied, in particular the Student-like shape of the
distribution of returns with a power-law tail index in the right range, and the slow decay of the volatility correlation
and of the kurtosis. The choice of the time scale $\tau$ is dictated by Figs. 4-5, and
is found to be on the order of $1/300$th of a day (a few minutes). The value of both $z_2$ and $\tau$ will probably
be affected by the inclusion of a non-Gaussian noise $\xi$ -- we leave the detailed study of this effect for future
work. 

\subsection{Volatility prediction}

There is another, more direct way to test the consistency of our model, 
which to some extent avoids the problem of the 
non-Gaussian nature of the noise $\xi$ 
(but is still confronted with the intrinsic problems of long memory and
non stationarity). The idea is to fix the exponent $\alpha$ and to regress, 
on empirical data, an estimate of
the daily square volatility on the computed feedback strength, defined as:
\be
X_i=\sum_{\ell=1}^\infty \frac{(x_i-x_{i-\ell})^2}{\ell^{1+\alpha}}.
\ee
In practice, we have estimated a noisy proxy of square volatility of a stock as 
$\sigma_i^2=(H-L+|O-C|)^2/4O^2$, 
where $O,H,L,C$ stand for Open, High, Low, Close. 
We have computed $X_i$ using the open prices and truncated 
the sum over $\ell$ beyond 500 days, which of course is not very accurate because when 
$\alpha$ is close to one, the above sum converges only very slowly. 

We then plot for all stocks $\sigma_i^2/\langle \sigma^2 \rangle$ vs. $X_i/
\langle \sigma^2 \rangle$. Using our data set, this gives $\sim 400,000$ points; the 
correlation coefficient between the
two sets is found to be $0.285$. This value is rather high in view of the roughness 
of our volatility proxy.  
The result is shown in Fig. 11, where we have performed a moving average over $1500$ points.
As one can see the assumption of a linear relation between $\sigma_i^2$ and $X_î$ 
is rather remarkably borne out,
over a rather large range of $X_i$. From the slope and intercept of the linear relation, 
we obtain a direct estimate of $z_2$, 
which we find to be $\approx 0.9$, quite close indeed to our previous determination. 
This direct estimate shows 
that (a) the basic assumption of the model, that past price changes feedback in the volatility 
as in Eq. (\ref{the-model}),
seems to be realistic; and (b) the model is indeed rather close to an instability, 
with a feedback mechanism that 
leads to a substantial increase of the volatility.

The direct determination of $\alpha$ using this method is however difficult: one could think 
of varying $\alpha$ and 
choosing the
value corresponding to the maximal correlation between $X_i$ and $\sigma_i$. 
Unfortunately, the dependence of this
correlation coefficient on $\alpha$ is weak and does not allow to extract a meaningful minimum, although one
can see that $\alpha=1.15$ is indeed in the range where the correlation is largest. 
One could also extend the above method to account for the leverage effect and estimate directly the 
asymmetry parameter $\varphi$.

\begin{figure}
\begin{center}
\psfig{file=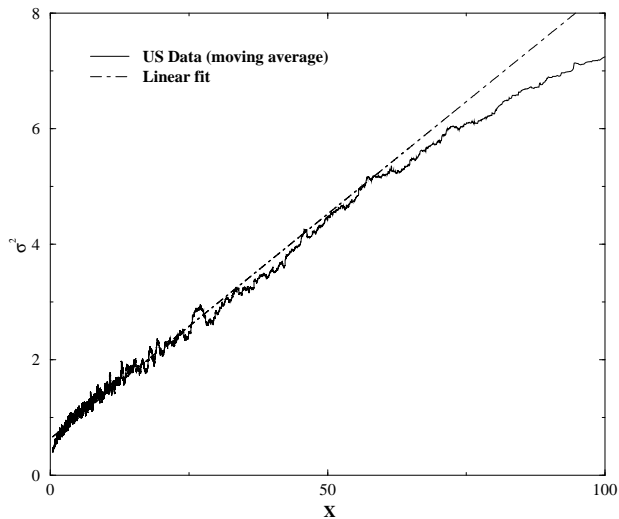,width=7cm,angle=270} 
\end{center}
\caption{Scatter plot of $\sigma_i^2$ vs. $X_i$ computed daily 
for 252 US stocks during the four-year period 2000-2003. The coordinates of 
each point were rescaled by the average square volatility of the stock during that time period. A moving average over
1500 points was performed, unveiling the nearly linear average behaviour of $\sigma_i^2$ on $X_i$, assumed in our
model. One can even notice a negative curvature for large $X$, as suggested by the saturation 
mechanism we invoked in section \ref{themodel} to motivate the model. 
}
\label{Fig11}
\end{figure}

\subsection{Summary}

In summary, we have shown that using a variety of different observables, 
the range of acceptable values for the 
parameters of the model can be approximately determined. We have found that
using these parameters, all stylized facts can be quantitatively accounted for.
Due to the proximity of the unstable regime, however, a very
precise determination of optimal parameters seems illusory.  
On the other hand, the basic assumption of our model,
that past price changes feedback in the volatility through Eq. (\ref{the-model}), 
is rather convincingly supported by the results shown in Fig. 11.

\section{Generalization to multiasset models}
\label{multiasset}

An interesting generalization of the above model concerns the multiasset situation, 
for example baskets of different
stocks with cross-correlations both in the returns and in the volatility. An obvious generalization of our model to this
case reads:
\be
x_{i+1}^a - x_i^a = r_i^a = \sigma_i^a \xi_i^a,
\ee
where $i$ denotes the time index and $a$ labels the stocks. The $\xi_i^a$'s are characterized by certain correlation
matrix $C_{ab} = \langle \xi_i^a \xi_i^b \rangle$ encoding the usual sectorial correlations. For the $\sigma_i^a$'s,
we write, in full generality:
\be\label{the-model-multiasset}
\sigma_i^{a2} = \sigma_0^{a2} \left[1 + \sum_{\ell=1}^\infty g_\ell \sum_{b} 
H^{ab} \frac{(x_i^b-x_{i-\ell}^b)}{\sigma_0^{b} \sqrt{\ell \tau}} + 
\sum_{\ell=1}^\infty g_\ell \sum_{b}  G^{ab} \frac{(x_i^b-x_{i-\ell}^b)^2}{\sigma_0^{b2} \ell \tau}\right].
\ee
We leave the investigation of this rich model for future work; thanks to the matrix 
structure of the feedback effect $H$ and $G$, one 
can reproduce a large variety of volatility cross-correlations and leverage effects. 
Here, we note that the average 
volatilities obey the 
following matrix equation:
\be
\sum_b \left(\delta_{ab} - \sum_{\ell=1}^\infty g_\ell G^{ab} \frac{\sigma_0^{a2}}{\sigma_0^{b2}}\right) 
\langle \sigma^{b2} \rangle = \sigma_0^{a2},
\ee
leading to a criterion for the stability of the model, which is that the smallest eigenvalue of the matrix 
on the left hand side of this equation must remain positive, generalizing the above criterion $1-z_2 < 1$. 

From Eq.(\ref{the-model-multiasset}) one can also estimate the leverage effect for index returns, which can be 
much enhanced if the matrix $H^{ab}$ has large off diagonal values compared to $G^{ab}$, meaning that a downward move on
any other stock $b$ is perceived as a source of risk for stock $a$, and triggers extra trades on all other stocks as well.

\section{Conclusion and perspectives}
\label{conclusion}

In this work, we have proposed and studied, both analytically and numerically, 
a multiscale feedback model of volatility.
This ARCH-like model (similar to the one studied by Zumbach in \cite{LZ}) 
assumes that the volatility is governed by the observed past price changes on different time scales,
which, we argue, directly influence the activity of traders. Assuming a power-law distribution 
of the time horizon of different traders, we obtain a model 
that captures most stylized facts of financial time 
series: Student-like distribution of returns with a power-law tail, long-memory of the volatility, slow convergence 
of the distribution
of returns towards the Gaussian distribution, multifractality and anomalous volatility relaxation after shocks. 
The model, at variance with recent multifractal models that are strictly time reversal invariant, reproduces the 
time asymmetry of financial time series revealed by Zumbach's mug-shots: past large scale volatility influence 
future small scale volatility.

The most important conclusion of our work is the following: in order to quantitatively reproduce empirical observations, 
the parameters must be chosen such that our model is `doubly' close to an instability, 
i.e. two parameters are close to values beyond which the process 
becomes 
non stationary. This means that (a) the feedback effect 
is important and substantially increases the volatility, and (b) that the model is intrinsically difficult to calibrate
because of the very long range nature of the correlations and the slow
convergence of all observables. However, by imposing the consistency of the model predictions 
with a large set of different empirical observations, a reasonable range of the parameters value can be determined. 
Furthermore, the adequacy of the basic assumption of our model, i.e. that the instantaneous 
volatility is directly related to 
a power-law superposition of past square returns on different time scales, can be directly assessed. The model can
easily be generalized to account for jumps (a feature needed to correct an unrealistic non monotonous behaviour
of the kurtosis), skewness and multiasset correlations. 

The interest of this type of models, compared to (multifractal) stochastic volatility models, is that their fundamental 
justification, in terms of agent based strategy, is relatively direct and plausible. We believe this is a strong constraint which should guide the construction of any mathematical model of reality. On the other hand, our 
fundamental assumption, Eq. (\ref{the-model}), is in contradiction with the efficient market hypothesis, which asserts 
that the price past history should have no bearing whatsoever on the behaviour of investors. The large correlation
that we find between past price changes and present volatility (see Fig. 11) indicates that this influence is in 
fact quite strong. This result is, in our view, yet another direct piece of evidence against the efficient 
market hypothesis, and a clear mechanism leading to excess volatility in financial markets.

Turning to financial engineering applications, such as risk control and option pricing, our model 
provides a well defined procedure to {\it filter} the series past price changes, and to compute
the probabilities of the different future paths. Similar models have been shown to fare rather well \cite{HARCH,Zumb2}. 
Once `softly' calibrated, the model can in principle be used for VaR estimates and option pricing. However, its mathematical 
complexity does not allow, in general, for explicit analytical solutions and probably one has to resort either to 
approximate treatments or to numerical, Monte-Carlo methods. The difficulty of long-memory models is that the option 
price must be computed conditional to the whole past history, which considerably complexifies both analytical 
solutions and Monte-Carlo methods. In other words, both the
option price and the optimal hedge are no longer simple functions of the current price, but {\it functionals} of the 
whole price history. Finding operational ways to account for this history dependence seems to us a major
challenge, on which we hope to work in the near future. 

\section*{Acknowledgments} We want to thank D. Farmer and T. Lux for inviting us to write this paper, which was initiated
by discussions during the Leiden workshop: ``Volatility in financial markets'', October 2004. 
Discussions with J.F. Muzy
and G. Zumbach have been of great help. L.B. also thanks J. Evnine for ongoing discussions and 
support.

\end{document}